\documentclass[aps,pra,twocolumn,showpacs,preprintnumbers,amsmath,amssymb]{revtex4-1}
\usepackage{graphicx}
\usepackage[usenames,dvipsnames]{xcolor}
\usepackage{hyperref}
\usepackage{cleveref}
\usepackage{makecell}
\usepackage{float}
\usepackage{lineno}
\usepackage[T1]{fontenc}
\usepackage[caption=false]{subfig}
\newcommand{\Tr}{\textrm{Tr}}

\newcommand{\bra}[1]{\ensuremath{\langle #1 |}}
\newcommand{\ket}[1]{\ensuremath{| #1 \rangle}}
\newcommand{\braket}[2]{\ensuremath{\langle #1 | #2 \rangle}}

\usepackage{amssymb,amsmath} 
\usepackage{amsfonts}
\usepackage{color}
\usepackage[utf8]{inputenc}
\usepackage{lmodern}
\usepackage{epsfig}
\usepackage{ulem}
\usepackage{blindtext}
\begin{document}

\title{Critical behavior of lattice gauge theory Rydberg simulators from effective Hamiltonians}
\author{Jin Zhang$^{1}$}
\email{jzhang91@cqu.edu.cn}
\author{S.-W. Tsai$^2$}
\author{Y. Meurice$^3$}
\affiliation{$^1$ Department of Physics and Chongqing Key Laboratory for Strongly Coupled Physics, Chongqing University, Chongqing 401331, China}
\affiliation{$^2$Department of Physics and Astronomy, University of California, Riverside, CA 92521, USA}
\affiliation{$^3$Department of Physics and Astronomy, University of Iowa, Iowa City, IA 52242, USA}
\definecolor{burnt}{cmyk}{0.2,0.8,1,0}
\def\lt{\lambda ^t}
\def\note{note}
\def\beq{\begin{equation}}
\def\enq{\end{equation}}

\date{\today}
\begin{abstract}
We consider multileg ladders of Rydberg atoms which have been proposed as quantum simulators for the compact Abelian Higgs model (CAHM) in 1+1 dimensions [Y. Meurice, Phys. Rev. D 104, 094513 (2021)] and modified versions of theses simulators such as triangular prisms. Starting with the physical Hamiltonian for the analog simulator, we construct translation-invariant effective Hamiltonians by integrating over the simulator high-energy states produced by the blockade mechanism when some of the atoms are sufficiently close to each others. Remarkably, for all the simulators considered, the effective Hamiltonians have the three types of terms present for the CAHM (Electric field, matter charge and currents energies) but, in addition, terms quartic in the electric field. For the two leg ladder, these additional terms cannot be removed by fine-tuning the adjustable parameters 
of currently available devices. For positive detuning, the new terms create highly-degenerate vacua resulting in a very interesting phase diagram. Using numerical methods, we demonstrate the close correspondence between the physical simulator and the effective description for the ground state energy and real-time evolution. We discuss the phase diagram at fixed geometry with variable Rabi frequency and detuning and show that a rich variety of phases can be reached with potential interest in the context of QCD at finite density. We illustrate how the effective description can be used to design simulators with desirable properties from the point of view of constructing hybrid event generators.
\end{abstract}

\maketitle

%%%%%%%%%%%%%%%%%%%%%%%%%%%%%%%%%%%%%%%%%%%%%%%%%%%%%%%%%%%%%%%
\section{Introduction}\label{sec:introduction}
%%%%%%%%%%%%%%%%%%%%%%%%%%%%%%%%%%%%%%%%%%%%%%%%%%%%%%%%%%%%%%%

The idea of using  quantum devices to study the real-time evolution of strongly interacting particles in high-energy and nuclear physics has gained considerable interest in recent years \cite{Dalmonte:2016alw,Banuls:2019bmf,Kasper:2020akk,Bauer:2022hpo,Halimeh:2023lid,DiMeglio:2023nsa}.
In this context, the possibility of building arrays of Rydberg 
atoms of significant size with adjustable geometry and external parameters \cite{51qubits,keesling2019,Browaeys:2020kzz} offer many new possibilities for analog simulations of lattice field theory models. In addition, publicly available interfaces \cite{quera,pasqal} allow users to configure arrays involving hundreds of Rydberg atoms and run their own experiments. 
This sets the path for extensive empirical exploration by scientists who don't have direct access to this type of facilities. 

This new technology has been used to propose simulations of spin and lattice gauge theory models \cite{Celi:2019lqy,Surace:2019dtp,Notarnicola:2019wzb,cara,Slagle:2021ene,Heitritter:2022jik,PhysRevResearch.5.023010,Yang:2020yer,zhou2022quantum,RhineZ2Gauge2023,HomeierRydDyn2023,Domanti:2023qht}.
One of the simplest model is the Abelian Higgs model \cite{prd92,Kuno:2016ipi,Gonzalez-Cuadra:2017lvz,Zhang:2018ufj,cara,Chanda:2021kie}. In 1+1 dimensions, and after elimination of the non-compact Brout-Englert-Higgs mode, the Hamiltonian reads
\begin{equation}
\label{eq:hahm}
\hat{H}_{CAHM} = D\sum_{i=1}^{N_{s}} \left(\hat{L}^z_{i}\right)^2 -Y{\sum_{i=1}^{N_s-1}}\hat{L}^z_{i} \hat{L}^z_{i+1}
	-X\sum_{i=1}^{N_{s}} \hat{U}^x_{i} \ 
 \end{equation}
with $N_s$ the number of sites, $\hat{L}^z\ket{m}=m\ket{m}$ and 
$m$ a discrete electric field quantum number ($-\infty< m<+\infty$), and 
$\hat{U}^x\equiv \frac{1}{2}(\hat{U}^+ + \hat{U}^-) $ with 
$\hat{U}^\pm\ket{m}=\ket{m\pm1}$. Note that in the derivation of the Hamiltonian we obtain a term of the form $\frac{Y}{2} \sum_i(\hat{L}^z_{i+1} - \hat{L}^z_{i})^2$ that accounts for matter interactions, however for matching purposes which become clear later, we have reabsorbed the local quadratic couplings in the definition of the coupling $D$. 
In practice, we also apply  truncations:  for a spin-$m_{max}$ truncation we have 
$\hat{U}^\pm\ket{\pm m_{max}}=0$. In the following we focus on the spin-1 truncation. 

It has been argued \cite{cara} that this spin-1 Hamiltonian can be approximately simulated using arrays of Rydberg atoms by adapting the optical lattice construction using $^{87}Rb$ atoms separated by controllable (but not too small) distances, coupled to the excited Rydberg state $\ket{r}$ with a detuning $\Delta$. 
The ground state is denoted $\ket{g}$ 
and the two possible states  $\ket{g}$ and $\ket{r}$ can be seen as a qubit with $n\ket{g}=0, \ n\ket{r}=\ket{r}$. 
The Hamiltonian for a generic array reads
\begin{equation}
\label{eq:genryd}
\hat{H} = \frac{\Omega}{2}\sum_i(\ket{g_i}\bra{r_i} + \ket{r_i}\bra{g_i})-\Delta\sum_i  \hat{n}_i +\sum_{i<j}V_{ij}\hat{n}_i\hat{n}_j,
\end{equation}
where the indices label atoms and 
$V_{ij}=\Omega R_b^6/r_{ij}^6$
for a distance $r_{ij}$ between the atoms labelled as $i$ and $j$. 
The Rydberg blockade radius $R_b$ is defined by the condition $V_{ij}=\Omega$ when $r_{ij}=R_b$. 

The simplest simulator is a two-leg ladder \cite{cara} with 
the correspondence:
\begin{equation}
\ket{gg} \rightarrow \ket{m=0},\ \ket{gr} \rightarrow \ket{m=1},\ \ket{rg} \rightarrow \ket{m=-1}, 
\end{equation}
for 
two atoms on a rung. If the distance separating these two atoms is small enough, the state $\ket{rr}$ is unlikely to appear.
In Ref. \cite{cara}, it was shown that a perfect matching could be obtained for the individual sites, however, the nearest-neighbor (NN) interactions controlled by $Y$ could not be matched exactly using the current technology where all the interactions are repulsive. 

In this article, we show that it is possible to construct an effective Hamiltonian for the simulator provided that the set of 
atoms used to emulate the local spin-1 degrees of freedom are close enough, in other words, if the distance separating them is less than $R_b$. We will show that if these sets form the rungs of a ladder and if the distance between the next-nearest-neighbor (NNN) rungs is larger than $R_b$, the effective Hamiltonian has the three types of terms found in Eq.~\eqref{eq:hahm} with in addition a quartic term of the form $\sum_{i=1}^{N_s-1} (\hat{L}^z_i)^2 (\hat{L}^z_{i+1})^2$. The ladder models considered and their effective Hamiltonians are presented  in Sec. \ref{sec:models}. 
Numerical tests are performed in Sec. \ref{sec:tests}. Their phase diagrams are discussed in Sec. \ref{sec:phases}. Practical applications
involving systems that can be simulated at facilities such as QuEra are discussed in Sec. \ref{sec:practical}. Implications for hybrid algorithms are briefly discussed in the conclusions.

%%%%%%%%%%%%%%%%%%%%%%%%%%%%%%%%%%%%%%%%%%%%%%%%%%%%%%
\section{Rydberg Simulator}
\label{sec:models}
%%%%%%%%%%%%%%%%%%%%%%%%%%%%%%%%%%%%%%%%%%%%%%%%%%%%%%
\begin{figure}[t!]
    \centering
    \includegraphics[width=0.45\textwidth]{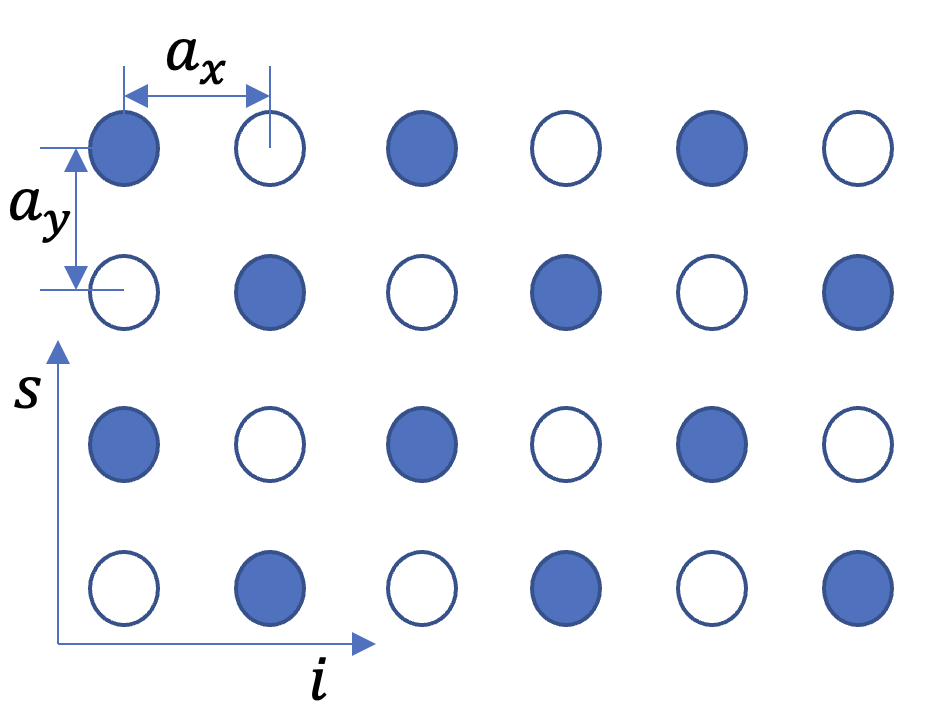}
    \caption{The rectangular multileg ladder of Rydberg atoms, $a_{x(y)}$ is the lattice spacing in $x(y)$ directions, $i$ labels rungs and $s$ labels legs. Here it is a checkerboard stripe, where empty circles represent atomic ground state $\ket{g}$, and the blue solid circles are excited Rydberg states $\ket{r}$.}
    \label{fig:4times6ladder}
\end{figure}

The multileg ladder of Rydberg atoms is a promising quantum simulator for exotic quantum critical phenomena and lattice gauge theories, as we can use different rung geometries to encode various local degrees of freedom. Figure~\ref{fig:4times6ladder} shows a rectangular multileg ladder of Rydberg atoms with $N_s$ rungs and $N_l$ legs. The Hamiltonian of the multileg ladder reads
\begin{eqnarray}
\label{eq:mLRHam}
\nonumber \hat{H}_{m\rm{LR}} &=& \frac{\Omega}{2} \sum_{\substack{i=1,2,\ldots,N_s \\ s=1,2,\ldots,N_l}} (\ket{g_{i,s}}\bra{r_{i,s}} + h.c.) - \Delta \sum_{\substack{i=1,2,\ldots,N_s \\ s=1,2,\ldots,N_l}} \hat{n}_{i,s} \\ &+& \sum_{(i,s) \neq (i',s')} V_{i,s;i',s'} \hat{n}_{i,s} \hat{n}_{i',s'},
\end{eqnarray}
where $\ket{r_{i,s}}$($\ket{g_{i,s}}$) is the Rydberg excited (ground) state at the site on the $i$th rung and the $s$th leg, $\hat{n}_{i,s} = \ket{r_{i,s}}\bra{r_{i,s}}$ is the Rydberg number operator, $\Omega$ and $\Delta$ are the Rabi frequency and detuning, respectively. The interaction between Rydberg states are long-range van der Waals repulsive interactions taking the form 
\begin{eqnarray}
\nonumber V_{i,s;i',s'} &=& \frac{C_6}{\left[(i-i')^2a_x^2+(s-s')^2a_y^2\right]^3} \\ &=& \frac{V_0}{\left[(i-i')^2/\rho^2+(s-s')^2\right]^3}
\end{eqnarray}
where $a_x$ ($a_y$) is the lattice spacing in $x$ ($y$) direction, $C_6$ is a constant, $V_0 = C_6/a_y^6$, and $\rho = a_y/a_x$ is the inverse aspect ratio. The Rydberg blockade mechanism means that at most one Rydberg state is allowed with a significant probability within a sufficiently small radius, typically the blockade radius $R_b$ that is defined by equating the interaction energy at distance $R_b$ to the Rabi frequency, $C_6/R_b^6 = \Omega$, thus $R_b = \left(C_6/\Omega \right)^{1/6}$. Below $R_b$, the interaction between Rydberg states is so strong that the laser field cannot excite two Rydberg states simultaneously \cite{RydPhy2018}. This imposes restrictions on the low-energy excitations for those building blocks of quantum systems. The Rydberg system is highly programmable. In our case, both the geometry of rungs and the parameters can be tuned to encode local spin degrees of freedom, and realize a class of spin chains. In the following, we label the leg by the spin projection quantum numbers.

%%%%%%%%%%%%%%%%%%%%%%%%%%%%%%%%%%%%%%%%
\subsection{Two-leg ladder}
%%%%%%%%%%%%%%%%%%%%%%%%%%%%%%%%%%%%%%%%
\begin{figure}[t!]
    \centering    \includegraphics[width=0.3\textwidth]{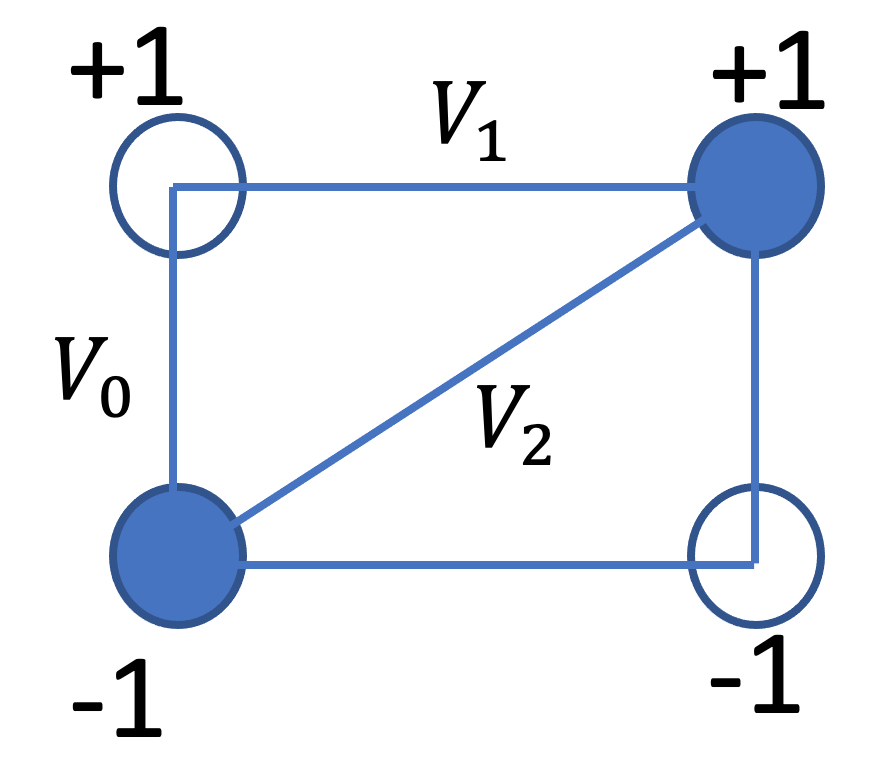}
    \caption{The two-leg Rydberg ladder with $N_s=2$ rungs. In each rung, the state with one Rydberg state in the upper leg is labeled by the spin $\ket{+1}$ state and the one with one Rydberg state in the lower leg is labeled by the spin $\ket{-1}$ state. The spin $\ket{0}$ state labels the state with no Rydberg state in the rung. The interactions between Rydberg states in the same rung, in the same leg, and in different legs are $V_0$, $V_1$, and $V_2$, respectively.}
    \label{fig:2legladder}
\end{figure}
\begin{figure}[t!]
    \centering
    \includegraphics[width=0.4\textwidth]{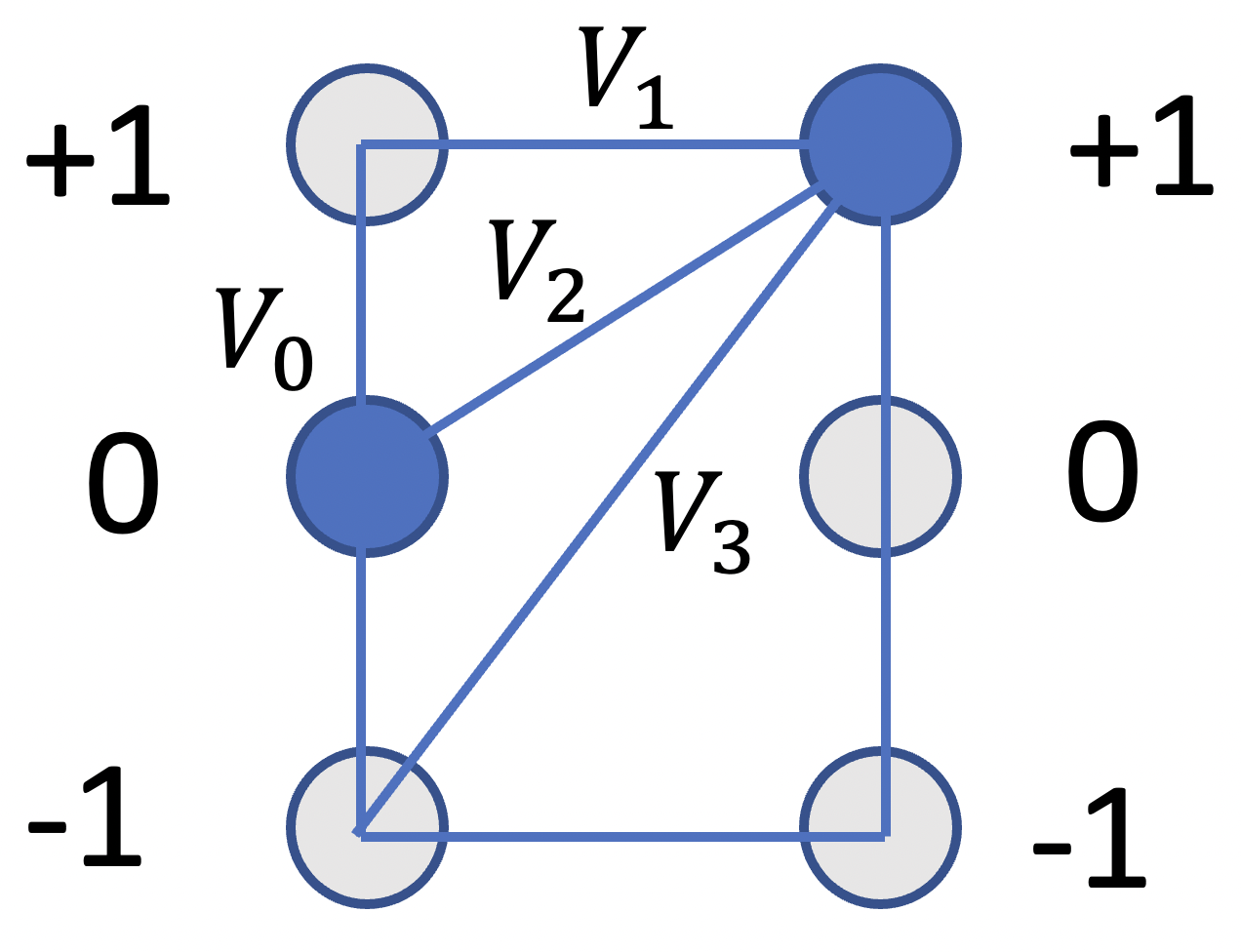}
    \caption{The three-leg Rydberg ladder with $N_s=2$ rungs. In each rung, the three states that have only one Rydberg state in the upper, middle, or lower leg are labeled by the spin $\ket{+1}$, $\ket{0}$, and $\ket{-1}$ states, respectively. In the same rung, the NN interaction is $V_0$, and the NNN interaction is $V'_0 = V_0/64$. The inter-rung interactions between atoms in the same leg, in NN legs, and in NNN legs are $V_1$, $V_2$, and $V_3$, respectively.}
    \label{fig:3legladder}
\end{figure}

The one-dimensional Rydberg chain is a special case of the multi-leg ladder system with $N_l = 1$. For a two-leg ladder, $N_l = 2$, there are four degrees of freedom in each rung: $\ket{g_{i,1}g_{i,2}}$, $\ket{g_{i,1}r_{i,2}}$, $\ket{r_{i,1}g_{i,2}}$, and $\ket{r_{i,1}r_{i,2}}$. The corresponding on-rung energy is $0$, $-\Delta$, $-\Delta$, and $V_0-2\Delta$, respectively. In principle, the four states can be used to represent the four projected states in $z$ direction for spin-$3/2$, then the on-rung interaction can be expressed by $\sum_{u=0}^{3} A_u (\hat{L}^z_{i})^u$, where the coefficients $A_u$ can be found by matching the energy spectrum. However, in real quantum systems, the most common onsite terms are the linear term that is coupled to the external field and the quadratic term that is the single-ion anisotropy. In addition, the Rydberg interaction is strong when the rung size is smaller than the Rydberg blockade radius such that the $\ket{r_{i,1}r_{i,2}}$ state is not likely to appear. Here we only consider a spin-$1$ realization by mapping the first three states to the spin-$1$ projected states $\ket{0}, \ket{+1}, \ket{-1}$, respectively. The onsite interaction term in spin language is thus $-\Delta \left(\hat{L}^z_i \right)^2$. The relation of the $z$-component spin operator to the Rydberg number operator is defined as \cite{cara}
\begin{equation}
\hat{L}^z_i = \hat{n}_{i,+1} - \hat{n}_{i,-1}.
\label{eq:number}
\end{equation}
If we take the square of this equation, use the property $\hat{n}^2=\hat{n}$, and drop the term $\hat{n}_{i,+1}\hat{n}_{i,-1}$ which is zero in the low energy sector we obtain effectively
\begin{equation}
    \left(\hat{L}^z_i\right)^2 = \hat{n}_{i,+1} + \hat{n}_{i,-1}
    \label{eq:oprelations1}
\end{equation}
Solving for $\hat{n}_{i,m}$ we get
\begin{eqnarray}
\label{eq:oprelations2}
\hat{n}_{i,+1(-1)} = \left[\left(\hat{L}^z_i \right)^2 \pm \hat{L}^z_i \right]/2.
\end{eqnarray}

Plugging Eq.~\ref{eq:oprelations2} into the Rydberg interactions, we can easily write down the interactions in terms of spin operators. If $\rho \ll 1$, we can just keep the nearest-neighbor-rung interactions because of fast decaying of the van der Waals interactions. Figure~\ref{fig:2legladder} shows the interactions between nearest-neighbor (NN) rungs, which contains the Rydberg interactions between atoms in the same leg $V_1 = V_0\rho^6$ and those between atoms in different legs $V_2 = V_0\rho^6/\left(1+\rho^2\right)^3$. The NN spin interactions are
\begin{equation}
\label{eq:spinnninteraction}
\hat{H}_{\rm 2LR,NN} = \frac{V_1-V_2}{2} \hat{L}^z_i \hat{L}^z_{i+1} + \frac{V_1+V_2 }{2} \left(\hat{L}^z_i\right)^2 (\hat{L}^z_{i+1})^2.
\
\end{equation}
For generic values of $\rho$, it is needed to include the long-range interactions. The interactions between the spin at site $i$ and that at site $i+k$ takes the same form as Eq. (~\ref{eq:spinnninteraction}) by replacing $V_1, V_2$ by $V_1^{(k)} = V_0 \rho^6/k^6, V_2^{(k)} = V_0 \rho^6 / \left(k^2 + \rho^2 \right)^3$.

Finally, noticing that the Rydberg Rabi term can flip the spin projections between $\ket{0}$ and $\ket{\pm 1}$, but there is no direct flipping channel between $\ket{+1}$ and $\ket{-1}$, the Rabi term is equivalent to the spin-1 ladder operator. In summary, if the rung size of the two-leg ladder is smaller than the Rydberg blockade radius, or $V_0 \gg \Delta, \Omega$, the two-leg Rydberg ladder is an effective spin-1 chain
\begin{eqnarray}
\label{eq:effectiveham2leg}
\nonumber \hat{H}_{\rm{2LR}}^{\rm{eff}} &=& -\Delta \sum_{i=1}^{N_s} \left(\hat{L}_i^z\right)^2 + \sum_k \left( \frac{V_1^{(k)} - V_2^{(k)}}{2} \sum_{i=1}^{N_s-k}\hat{L}^z_i \hat{L}^z_{i+k} \right. \\ \nonumber && \left. + \frac{V_1^{(k)}+V_2^{(k)}}{2} \sum_{i=1}^{N_s-k} \left(\hat{L}^z_i\right)^2 \left(\hat{L}^z_{i+k}\right)^2 \right) \\ &&+ \frac{\Omega}{2} \sum_{i=1}^{N_s} \left( \hat{U}^+_i + \hat{U}^-_i \right).
\end{eqnarray}

We now consider the case where the long-range interactions have a negligible effect and keep only the NN interactions. In this situation the effective Hamiltonian reads
\begin{eqnarray}
\label{eq:effectiveham2legNN}
\nonumber \hat{H}^{\rm{eff}}_{\rm{2LR}} &=& -\Delta \sum_{i=1}^{N_s} \left(\hat{L}_i^z\right)^2 + \frac{V_1 - V_2}{2} \sum_{i=1}^{N_s-1}\hat{L}^z_i \hat{L}^z_{i+1} \\ \nonumber &&+ \frac{V_1+V_2}{2} \sum_{i=1}^{N_s-1} \left(\hat{L}^z_i\right)^2 \left(\hat{L}^z_{i+1}\right)^2 \\ &&+ \frac{\Omega}{2} \sum_{i=1}^{N_s} \left( \hat{U}^+_i + \hat{U}^-_i \right).
\end{eqnarray}
The matching with the target model requires 
\begin{itemize}
\item
$\Delta=-D$, note that the sign matters.
\item
The coefficient for $\hat{L}^z_i \hat{L}^z_{i+1}$ is positive for the simulator (repulsive/antiferromagnetic) but  the CAHM has ferromagnetic interactions. This can be remedied by redefining the observable 
$\hat{L}^z_{2i+1}\rightarrow - \hat{L}^z_{2i+1}$ (staggered interpretation).
\item
After this redefinition $V_1=-V_2=Y>0$ but $V_2>0$ with current technology.
\item
$\Omega=-X$, the sign does not matter because we can redefine the relative phase between $\ket{g}$ and $\ket{r}$ without physical consequences. 
\end{itemize}
These results agree with two-rung results of \cite{cara}.

%%%%%%%%%%%%%%%%%%%%%%%%%%%%%%%%%%%%%%%%
\subsection{Three-leg ladder}
%%%%%%%%%%%%%%%%%%%%%%%%%%%%%%%%%%%%%%%%

Another scheme to realize spin-one chains is to use a three-leg ladder, where the three states each with only one Rydberg state in a rung represent  the three spin projection states in $z$ direction \cite{cara}. The configuration is shown in Fig. \ref{fig:3legladder}, where five different types of interactions $V_0$, $V_0'$, $V_1$, $V_2$, and $V_3$ exist, and we allow a small offset for the detuning in the middle leg, $\Delta+\Delta_0$. There are eight states in a rung, four of which have more than one Rydberg states, and one of which has no Rydberg state. We consider two cases here. 

In case $1$, the whole rung is within the Rydberg blockade radius $2a_y < R_b$, only the four states with less than two Rydberg states are allowed in the low energy band. We can further tune the detuning $\Delta$ such that only the three states with one Rydberg state are allowed in the low-energy band. In case 1, we require $V_0, V'_0 \gg \Delta \gg |\Delta_0|, |\Omega|$. 

In case $2$, Rydberg states in NN legs in the same rung are blockaded by letting $a_y < R_b$, but the size of the whole rung is larger than $R_b$, and the state with two Rydberg states in the upper and the lower legs in the same rung may has a lower energy than those states with one Rydberg state, that is $-2\Delta + V_0' < -\Delta$ if $V_0'=V_0/64 < \Delta$. Thus the spin-$1$ sector is not in the lowest energy band. If the energy gap $\Delta - V_0' \gg \Omega$, the tunneling from the spin-1 sector to the ground state is small and the $3$-leg Rydberg system can still simulate the spin-$1$ dynamics with good accuracy. In case 2, we require $V_0, \Delta, |V_0-\Delta| \gg V'_0, |\Delta_0|, |\Omega|$. 

For both cases, the three spin projection states are nearly degenerate and form an energy band. They are nearly orthogonal to other Rydberg states.
In both cases, in order to obtain the effective spin-$1$ Hamiltonian, we adapt the relation between the $z$-component of the spin-$1$ operator and the Rydberg number operator \cite{cara}, proceed as in Eq.  (\ref{eq:oprelations1}), impose the constraint that there is exactly one Rydberg state for the three spin-1 states, and obtain the equations
\begin{eqnarray}
\nonumber \hat{L}^z_i &=& \hat{n}_{i,+1} - \hat{n}_{i,-1} \\ 
    \left(\hat{L}^z_i\right)^2 &=& \hat{n}_{i,+1} + \hat{n}_{i,-1} \\  \nonumber
    \hat{n}_{i,-1} &+& \hat{n}_{i,0} + \hat{n}_{i,+1} = 1\\ \nonumber
\end{eqnarray}

When the detuning term is a uniform constant $-\Delta$, the constraint among the occupation numbers in a rung implies that the detuning energy is 
a constant. If we allow a small offset $\Delta_0$ to the detuning in the middle leg \cite{cara}, the detuning energy $\hat{H}_{\rm{detun.}, i}$ at rung $i$ reads
\begin{equation}
\hat{H}_{\rm{detun.}, i} =\Delta_0 (\hat{L}^z_i)^2 - \Delta - \Delta_0.
\end{equation}
The interaction term between NN rungs $\hat{H}_{\rm{3LR, NN}}$ can be rewritten as
\begin{eqnarray} \nonumber
 \hat{H}_{\rm{3LR, NN}}&=& \left[\left( 3V_1+V_3 \right)/2 - 2V_2\right] \left(\hat{L}^z_i \right)^2 \left(\hat{L}^z_{i+1} \right)^2\\ &+& \left[\left(V_1-V_3\right)/2\right] \hat{L}^z_i \hat{L}^z_{i+1}\\ \nonumber &+& \left(V_2-V_1\right)\left[\left(\hat{L}^z_i\right)^2+\left(\hat{L}^z_{i+1}\right)^2\right] + V_1 . 
\end{eqnarray}

The effective Hamiltonian for the Rabi term is different for the two cases mentioned above. In case $1$, the whole rung is blockaded, the Rabi term can flip between each two of the three spin states via hopping back and forth between the empty state and the spin-$1$ sector. Notice that there is also possibility for the atom jumping out from and going back to the same state, so the Rabi term effectively contributes both a diagonal term and a clock ladder operator $\hat{C}^{\pm}_i$ which provides (anti)cyclic permutations.

In case $2$, the energy of the $\ket{rgr}$ state is about $|\Delta|$ lower than the states in the spin-$1$ sector ($V_0'$ is small), while the energy of the empty state is the same amount higher. The hopping amplitudes of flipping between $\ket{+1}$ and $\ket{-1}$ via the empty state and that via the $\ket{rgr}$ state have the nearly the same magnitude but opposite signs so are canceled. The Rabi term is effectively a spin-$1$ ladder operator $\hat{U}^{\pm}_i$. 

We can come to the same conclusion using perturbation theory to calculate the effective coupling constants. Treating the Rabi term as a perturbation, the effective Hamiltonian for the Rabi term is
\begin{eqnarray}
\label{eq:ptmatrix}
\nonumber \bra{e'}\hat{H}^{\rm{eff}}_{\rm{3LR}, \Omega}\ket{e} &=& \frac{1}{2}\sum_{f \neq e, e'} \bra{e'}\hat{H}_{\rm{3LR}, \Omega}\ket{f}\bra{f}\hat{H}_{\rm{3LR}, \Omega}\ket{e} \\ && \left(\frac{1}{E_{e'} - E_f}-\frac{1}{E_f-E_e}\right),
\end{eqnarray}
where $\ket{e}$($\ket{f}$) is the state inside(outside) of the spin-1 sector. The effective matrix for the Rabi term in the spin-$1$ sector is
\begin{eqnarray}
\label{eq:effrabi}
- \frac{\Omega^2}{4}
\begin{pmatrix}
 \mathrm{A} & \Gamma & \Lambda\\

\Gamma & \mathrm{B} & \Gamma \\

\Lambda & \Gamma & \mathrm{A}
\end{pmatrix}
\end{eqnarray}
where
\begin{eqnarray}
\nonumber \mathrm{A} &=& \frac{1}{V_0 - \Delta-\Delta_0}+\frac{1}{V_0'-\Delta} + \frac{1}{\Delta} \\ \nonumber 
\mathrm{B} &=& \frac{2}{V_0-\Delta}+\frac{1}{\Delta+\Delta_0} \\ \nonumber 
2\Gamma &=& \frac{1}{\Delta}+\frac{1}{V_0-\Delta}+\frac{1}{\Delta+\Delta_0}+\frac{1}{V_0-\Delta-\Delta_0} \\
\Lambda &=& \frac{1}{V'_0-\Delta}+\frac{1}{\Delta}
\end{eqnarray}
The diagonal part of the effective matrix can be written as $[(\mathrm{B}-\mathrm{A})\Omega^2/4] (\hat{L}^z_i)^2$ up to a constant. For case $1$, $\Gamma \approx \Lambda \approx 1/\Delta$, the off-diagonal part is a three-state clock ladder operator. For case $2$, $\Lambda \approx 0$, the off-diagonal part is a spin-$1$ ladder operator. Thus effectively
\begin{eqnarray}
    \hat{H}^{\rm{eff}}_{\rm{3LR},\Omega} = \begin{cases}
    -\frac{\Omega^2}{4\Delta} \sum_i \left(\hat{C}_i^+ + \hat{C}_i^-\right) & \text{for case 1,} \\
    -\frac{\Omega^2\Gamma}{4} \sum_i \left(\hat{U}^+_i + \hat{U}^-_i\right)  & \text{for case 2.}
\end{cases}
\end{eqnarray}

In summary, one can write down the effective spin-$1$ Hamiltonian for the three-leg Rydberg ladder
\begin{eqnarray}
\label{eq:effectiveham3legNN}
\nonumber \hat{H}^{\rm{eff}}_{\rm{3LR}} &=& \left[\Delta_0 + \frac{\Omega^2\left(\mathrm{B}-\mathrm{A}\right)}{4}\right] \sum_{i=1}^{N_s} \left(\hat{L}_i^z\right)^2 \\ \nonumber &+& \left(V_2 - V_1 \right) \sum_{i=1}^{N_s-1} \left[\left(\hat{L}^z_i\right)^2 + \left(\hat{L}^z_{i+1}\right)^2 \right] \\ \nonumber &+& \frac{V_1-V_3}{2} \sum_{i=1}^{N_s-1}\hat{L}^z_i \hat{L}^z_{i+1} \\ \nonumber &+& \left(\frac{3V_1+V_3}{2} - 2V_2 \right) \sum_{i=1}^{N_s-1} \left(\hat{L}^z_i\right)^2 \left(\hat{L}^z_{i+1}\right)^2 \\ &+& \hat{H}^{\rm{eff}}_{\rm{3LR}, \Omega} + \rm{Const.} \hat{I},
\end{eqnarray}
where the constant $\rm{Const.} = -(\Delta+\Delta_0)N_s - N_s\Omega^2B/4 + V_1(N_s-1)$.

\begin{figure}[t!]
    \centering
    \includegraphics[width=0.45\textwidth]{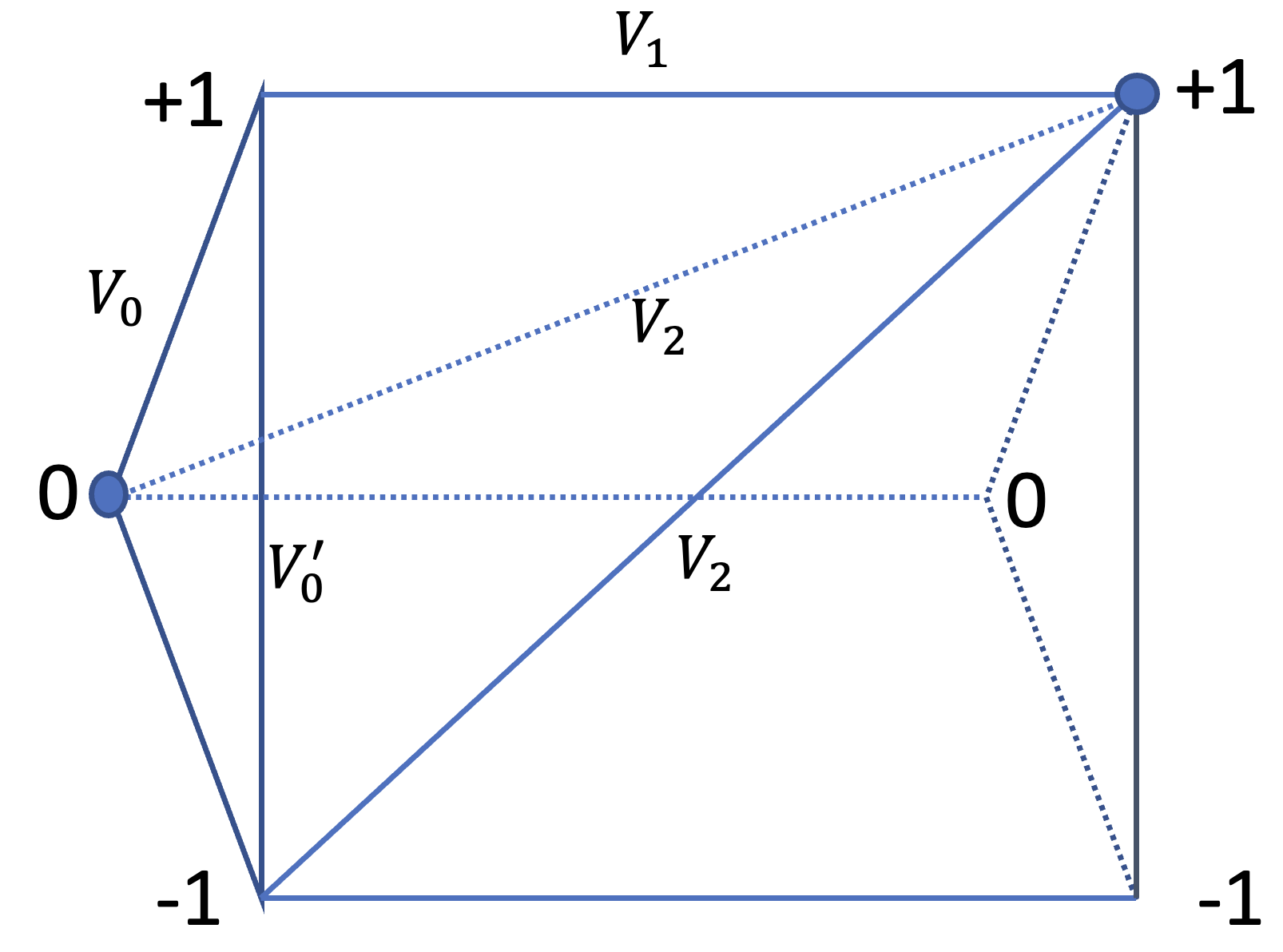}
    \caption{Same as Fig.~\ref{fig:3legladder}, but the middle leg is put out of plane to form a equilateral triangular prism. }
    \label{fig:triangularprism}
\end{figure}
\begin{figure}[t!]
    \centering
    \includegraphics[width=0.48\textwidth]{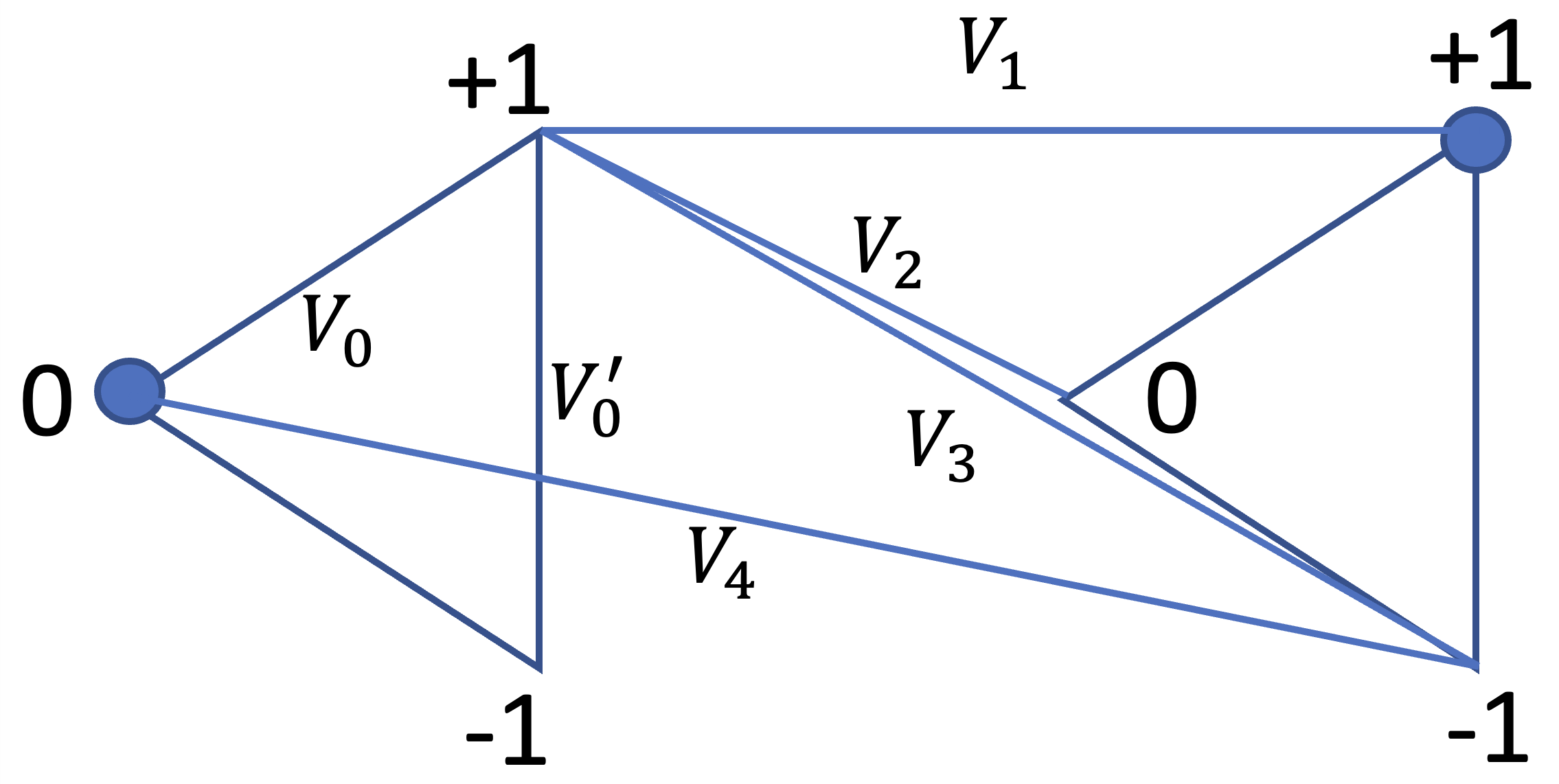}
    \caption{Same as Fig.~\ref{fig:triangularprism}, but triangles reside in the same plane.}
    \label{fig:trianglesinplane}
\end{figure}
%%%%%%%%%%%%%%%%%%%%%%%%%%%%%%%%%%%%%%%%
\subsection{Triangular prism}
\label{subsec:triangularprism}
%%%%%%%%%%%%%%%%%%%%%%%%%%%%%%%%%%%%%%%%

If we allow the middle leg of the three-leg ladder moving out of plane, the effective Hamiltonian for case 1 does not change as long as the size of the whole rung is within $R_b$. In particular, if the three sites in a rung form an equilateral triangle (Fig.~\ref{fig:triangularprism}), $V_2 = V_3$, and $V_0 = V'_0 \gg \Delta \gg |\Delta_0|,|\Omega|$. Thus $\Gamma \approx \Lambda$, $B \approx A$, the effective Hamiltonian for the Rydberg simulator is
\begin{eqnarray}
\label{eq:effectivehamc1}
\nonumber \hat{H}^{\rm{eff}}_{\rm{prism}} &=& \Delta_0 \sum_{i=1}^{N_s} \left(\hat{L}_i^z\right)^2 + \left(V_2 - V_1 \right) \sum_{i=1}^{N_s-1} \left[\left(\hat{L}^z_i\right)^2 + \left(\hat{L}^z_{i+1}\right)^2 \right] \\ \nonumber &+& \frac{V_1-V_2}{2} \sum_{i=1}^{N_s-1}\hat{L}^z_i \hat{L}^z_{i+1} \\ \nonumber &+& \frac{3(V_1-V_2)}{2} \sum_{i=1}^{N_s-1} \left(\hat{L}^z_i\right)^2 \left(\hat{L}^z_{i+1}\right)^2 \\ &-& \frac{\Omega^2V_0}{4\Delta(V_0-\Delta)} \sum_{i=1}^{N_s} \left( \hat{C}^+_i + \hat{C}^-_i \right) + \rm{Const.} \hat{I}.
\end{eqnarray}
Here, the coefficient of $\hat{L}^z_{i}\hat{L}^z_{i+1}$ and that of $(\hat{L}^z_{i})^2(\hat{L}^z_{i+1})^2$ have fixed ratio, the physics originated from the competition between the two terms may be missing in the Hamiltonian. But the blockade radius required is just half of the one in Fig.~\ref{fig:3legladder}.

%%%%%%%%%%%%%%%%%%%%%%%%%%%%%%%%%%%%%%%%
\subsection{Triangles in the same plane}
\label{subsec:triangleinplane}
%%%%%%%%%%%%%%%%%%%%%%%%%%%%%%%%%%%%%%%%

As the three dimensional triangular prism is not easy to realize in experiment, we can just shift the middle leg in the same plane. As shown in Fig.~\ref{fig:trianglesinplane}, we shift the middle leg to left (or right). Now the off-rung interactions between nearest-neighbor legs depend on its location and take two values $V_2, V_4$. The effective Hamiltonian reads
\begin{eqnarray}
\label{eq:effectivehaminplanec1}
\nonumber \hat{H}^{\rm{eff}}_{\rm{in-plane}} &=& \Delta_0 \sum_{i=1}^{N_s} \left(\hat{L}_i^z\right)^2 \\ \nonumber &+& \sum_{i=1}^{N_s-1} \left[\left(V_2 - V_1 \right)\left(\hat{L}^z_i\right)^2 + \left(V_4 - V_1 \right) \left(\hat{L}^z_{i+1}\right)^2 \right] \\ \nonumber &+& \frac{V_1-V_3}{2} \sum_{i=1}^{N_s-1}\hat{L}^z_i \hat{L}^z_{i+1} \\ \nonumber &+& \left(\frac{3V_1+V_3}{2} - V_2 - V_4 \right) \sum_{i=1}^{N_s-1} \left(\hat{L}^z_i\right)^2 \left(\hat{L}^z_{i+1}\right)^2 \\ &-& \frac{\Omega^2V_0}{4\Delta(V_0-\Delta)} \sum_{i=1}^{N_s} \left( \hat{C}^+_i + \hat{C}^-_i \right) + \rm{Const.} \hat{I}.
\end{eqnarray}
By shifting the middle leg, we have more degrees of freedom to tune the coefficients of the interaction terms: if $V_0$, $\Delta$ and $\Omega$ are fixed, we can tune $\Delta_0$, $a_x$ and the angle of the triangle to tune couplings of $(\hat{L}^z_i)^2$,  $\hat{L}^z_i\hat{L}^z_{i+1}$, and $(\hat{L}^z_i)^2(\hat{L}^z_{i+1})^2$ terms.

%%%%%%%%%%%%%%%%%%%%%%%%%%%%%%%%%%%%%
\section{Numerical tests}
\label{sec:tests}
%%%%%%%%%%%%%%%%%%%%%%%%%%%%%%%%%%%%%
\begin{figure}[t!]
    \centering
    \includegraphics[width=0.49\textwidth]{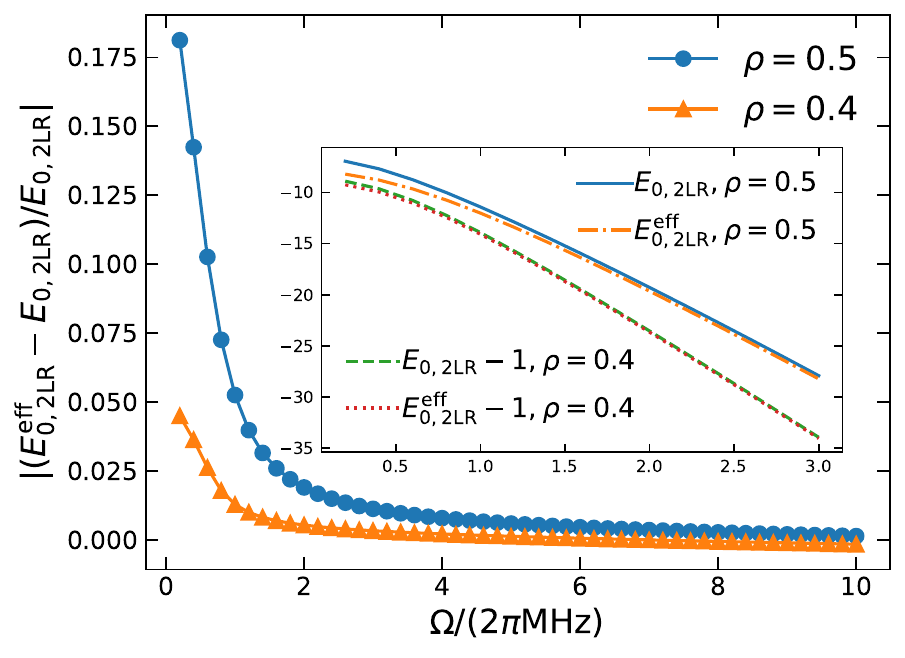}
    \caption{The relative error of the ground-state energy $E^{\rm{eff}}_{0,\rm{2LR}}$ of the effective Hamiltonian~\eqref{eq:effectiveham2legNN} as a function of $\Omega$. All the van der Waals interactions are kept in the  original two-leg Rydberg ladder to compute the actual ground-state energy $E_{0,\rm{2LR}}$. Here, $N_s = 16$, $V_0 = 1000\times 2\pi$ MHz, $\Delta = 1\times 2\pi$ MHz, and two inverse aspect ratios $\rho = 0.5, 0.4$ are considered. The inset shows the ground-state energy as a function of $\Omega$. The energy for $\rho=0.4$ is shifted by $-1$ for better view.}
    \label{fig:gse0vsomega2leg}
\end{figure}
We have derived the low-energy effective Hamiltonian for various Rydberg ladder systems using perturbation theory. All the effective spin-1 Hamiltonians obtained in Sec.~\ref{sec:models} have the same form
\begin{eqnarray}
\label{eq:effectivehamgeneral}
\nonumber \hat{H}^{\rm{eff}} &&= D \sum_{i=1}^{N_s} \left(\hat{L}_i^z\right)^2 + R \sum_{i=1}^{N_s-1}\hat{L}^z_i \hat{L}^z_{i+1}\\
 +&& R' \!\! \sum_{i=1}^{N_s-1} \!\!\left(\hat{L}^z_i\right)^2 \left(\hat{L}^z_{i+1}\right)^2 \!-\! J\! \sum_{i=1}^{N_s} \left( \hat{U}^+_i + \hat{U}^-_i \right),
\end{eqnarray}
where the ladder operator $\hat{U}^{\pm}_i$ can be replaced by the clock raising and lowering operator $\hat{C}^{\pm}_i$ for the cases that allow hopping between $\ket{\pm1}$ states. In this section, we provide some numerical calculations to test the validity of our results. The numerical results for large system sizes are calculated by the density matrix renormalization group (DMRG) algorithm.  Our DMRG calculations are performed with \textsc{ITensor Julia Library} \cite{10.21468/SciPostPhysCodeb.4}. When searching for the ground state, we gradually increase the maximum bond dimension during the variational sweeps until the truncation error $\epsilon$ is below $10^{-10}$. DMRG sweeps are terminated once the ground-state energy changes less than $10^{-11}$ and the von Neumann entanglement entropy changes less than $10^{-8}$ between the last two sweeps.

%%%%%%%%%%%%%%%%%%%%%%%%%%%%%%%%%
\subsection{Two-leg ladder}
%%%%%%%%%%%%%%%%%%%%%%%%%%%%%%%%%
In this case, the effective Hamiltonian is shown in Eq.~\eqref{eq:effectiveham2legNN} with $D = -\Delta$, $R = (V_2-V_1)/2$, $R' = (V_1+V_2)/2$, $J = -\Omega/2$. Since the error of the effective Hamiltonian is of order $\Omega^2/(4V_0)$, we set $V_0$ to a large value. Given that $V_{ij}=C_6/r_{ij}^6$ with $C_6 = 858386 \times 2\pi$ MHz $\mu$m$^6$, we set $V_0=1000\times 2\pi$ MHz by taking the length of the rung to be 
$a_y = 3.083 \mu$m. Figure~\ref{fig:gse0vsomega2leg} presents the ground-state energy difference between the two-leg Rydberg ladder $\hat{H}_{\rm{2LR}}$ and the corresponding effective Hamiltonian~\eqref{eq:effectiveham2legNN} as a function of $\Omega$ for $N_s = 16$, $\Delta=1\times 2\pi$ MHz. The effective Hamiltonian only contain NN interactions, while $\hat{H}_{\rm{2LR}}$ contains all van der Waals interactions. If the inverse aspect ratio $\rho=0.5$, the relative energy difference is about $18\%$ for small $\Omega/(2\pi \rm{MHz})=0.2$, and decreases quickly toward zero as $\Omega$ is increased. For smaller inverse aspect ratio $\rho=0.4$, the relative difference behaves the same way but with an overall smaller magnitude. The inset shows that the absolute energy difference also decreases with increasing $\Omega$ and decreasing $\rho$.

Note that the error from perturbation theory is of order $\Omega^2/(4V_0)$, which should increase as we increase $\Omega$, but it is small compared to the error from the omitted long-range interactions. For small $\Omega$, the blockade radius $R_b/a_y = (V_0/\Omega)^{1/6}$ is large and the interactions beyond NN rungs play important roles and are not negligible, so the relative energy difference is large. As we increase $\Omega$, the blockade radius $R_b$ is decreased and the effects of long-range interactions fade away, so the relative energy difference is small. By fixing $a_y$ and increasing $a_x$ such that $\rho$ is decreased, $R_b$ is becoming relatively smaller compared to $a_x$, so the effects of long-range interactions also decrease. For $1 < \Omega/(2\pi \text{MHz}) < 10$, $2.154 < R_b/a_y < 3.162$, the Rydberg blockade radius is smaller than $2a_x$ and two Rydberg states are allowed in NNN rungs 
for both $\rho=0.5$ and $\rho=0.4$. One can see that the relative energy difference is below $5\%$ for $1 < \Omega/(2\pi \text{MHz}) < 10$. We conclude that it is safe to omit interactions with a range longer than NN rungs when the blockade radius $R_b$ is smaller than the NNN distance.

%%%%%%%%%%%%%%%%%%%%%%%%%%%%%%%%%
\subsection{Three-leg ladder}
%%%%%%%%%%%%%%%%%%%%%%%%%%%%%%%%%
\begin{figure}[t!]
    \centering
    \includegraphics[width=0.5\textwidth]{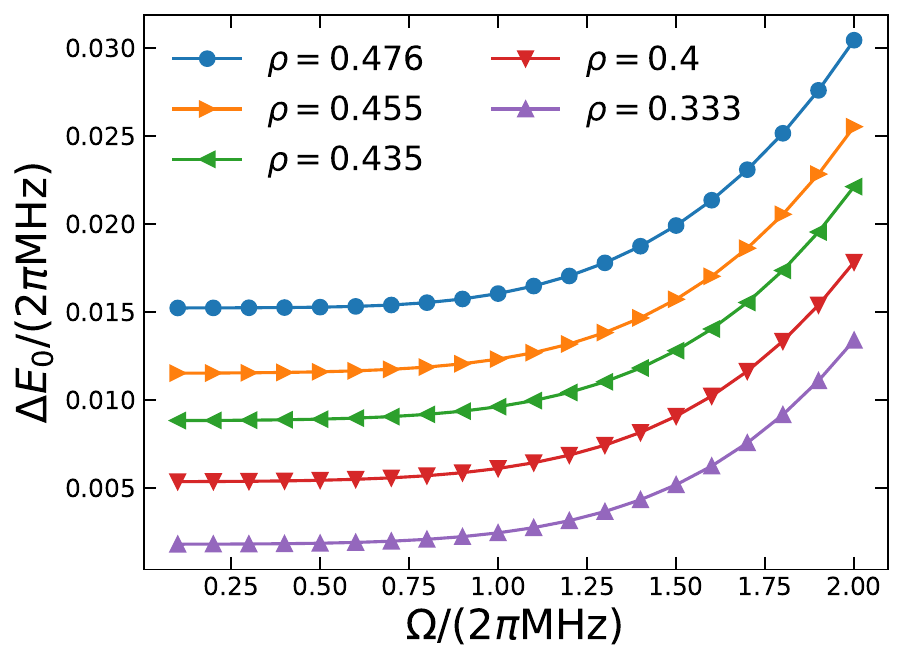}
    \caption{The ground-state energy difference between the effective Hamiltonian in Eq.~\eqref{eq:effectiveham3legNN} for case 2 and the real three-leg Rydberg ladder. The system size is $N_s = 4$ and the parameters are $V_0 = 2\Delta = 40 \times 2\pi \rm{MHz}$, and $\Delta_0 = 0.5\times 2\pi \rm{MHz}$. Five values of the inverse aspect ratio $\rho = a_y/a_x$ are considered here.}
    \label{fig:3legladdercompareE}
\end{figure}

The effective Hamiltonian for the three-leg Rydberg ladder is shown in Eq.~\eqref{eq:effectiveham3legNN}, where there are two limiting cases with different quantum operators $\hat{C}_i^{\pm}$ or $\hat{U}_i^{\pm}$. Because the error from the perturbation theory is well controlled, we just consider case $2$. Matching the form in Eq.~\eqref{eq:effectivehamgeneral}, we have $D = \Delta_0 + \Omega^2(B-A)/4 + 2(V_2-V_1)$, $R = (V_1-V_3)/2$, $R' = (3V_1+V_3)/2-2V_2$, $J = \Omega^2\Gamma/4$. Notice that on both ends $D_1 = D_{N_s} = \Delta_0 + \Omega^2(B-A)/4 + (V_2-V_1)$, but the boundary terms will not affect the bulk properties in the thermodynamic limit. We can use exact diagonalization to show that the real three-leg Rydberg ladder system and the effective Hamiltonian have close energy for small system sizes. The results for $N_s = 4$, $V_0 = 2\Delta = 40\times 2\pi$ MHz, $\Delta_0 = 0.5\times 2\pi$ MHz with different inverse aspect ratios are presented in Fig.~\ref{fig:3legladdercompareE}. One can see that the ground-state energy difference between $\hat{H}_{\rm{3LR}}$ and $\hat{H}^{\rm{eff}}_{\rm{3LR}}$ is small for all the inverse aspect ratios ($\rho$) considered here. The energy difference decreases as $\rho$ increases, because the effects of long-range interactions become small as $a_x$ increases. When $\Omega$ increases, the error from the second-order perturbation theory increases, so as the energy difference.

\begin{figure}[t!]
    \centering
    \includegraphics[width=0.5\textwidth]{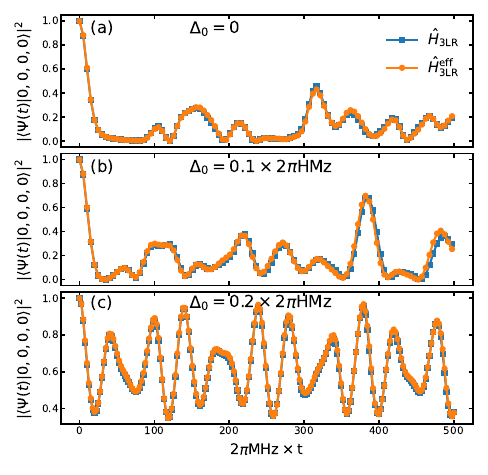}
    \caption{The real-time evolution of the initial state $\ket{0,0,0,0}$ with the effective Hamiltonian~\eqref{eq:effectiveham3legNN} for case 2 ($\hat{H}^{\rm{eff}}_{\rm{3LR}}$) and that with the three-leg Rydberg ladder Hamiltonian ($\hat{H}_{\rm{3LR}}$). Here, $N_s = 4$, $V_0 = 2\Delta = 40\times 2\pi \rm{MHz}$, $\Omega = 1\times 2\pi \rm{MHz}$, and $\rho = 0.4$. Three values of $\Delta_0/(2\pi \rm{MHz}) = 0, 0.1$, and $0.2$ are considered for (a), (b), and (c), respectively.}
    \label{fig:timeevolve3legladder}
\end{figure}
For large system sizes, one expects to use DMRG to calculate the energy. However, because the spin-1 sector is not the lowest energy band in the real three-leg Rydberg ladder, the low-energy states in the effective Hamiltonian correspond to the high-energy states in the real Rydberg system. We cannot check the effective Hamiltonian using normal DMRG which finds the ground state. We have seen that the effective Hamiltonian for the two-leg Rydberg ladder works well, so we believe the effective Hamiltonian in Eq.~\eqref{eq:effectiveham3legNN} also well describe the physics of the spin-1 sector. If one energy band largely separates from other bands, the time-evolved quantum state from an initial state in that band will stay inside the band for a long time. The real three-leg Rydberg ladder can still simulate the dynamics of the effective spin Hamiltonian with high precision.

We calculate the time evolution of the initial state with all the $L^z_i = 0$ for the effective Hamiltonian and compare it with the time evolution of the initial state with all the rungs having only one Rydberg state in the middle leg for the real Rydberg system. The results are shown in Fig.~\ref{fig:timeevolve3legladder} for $N_s = 4$, $V_0 = 2\Delta = 40\times 2\pi$ MHz, $\Omega=1\times 2\pi$ MHz, and $\rho = 0.4$. Then $R/(2\pi \rm{MHz}) \approx 0.06335$, $R'/(2\pi \rm{MHz}) \approx 0.05440$, and $J/(2\pi \rm{MHz}) \approx 0.02500$. Three values of $\Delta_0/(2\pi \rm{MHz}) = 0, 0.1$, and $0.2$ are considered, which corresponds to $D/(2\pi \rm{MHz}) = -0.092346, 0.007529$, and $0.107404$, respectively. It is seen that the probability of detecting the initial state $\ket{0,0,0,0}$ in the time-evolved quantum state $\ket{\Psi(t)}$, $|\braket{\Psi(t)}{0,0,0,0}|^2$, for the effective Hamiltonian is almost the same as that for the real three-leg Rydberg ladder. This good agreement persists for a long time $> 79.6 \rm{\mu s}$ as shown in Fig.~\ref{fig:timeevolve3legladder}. We remark that the values of $D, R, R'$, and $J$ are in the same order here and the system is not in the limiting case of the weak interacting regime. Therefore, our results confirm that the derived effective Hamiltonian in Eq.~\eqref{eq:effectiveham3legNN} accurately describes the many-body physics of the spin-1 sector for the real three-leg Rydberg ladder. In other words, the three-leg Rydberg ladder can quantum simulate the many-body dynamics of the effective spin-1 Hamiltonian in Eq.~\eqref{eq:effectiveham3legNN}.

%%%%%%%%%%%%%%%%%%%%%%%%%%%%%%%%%
\subsection{Other geometries}
%%%%%%%%%%%%%%%%%%%%%%%%%%%%%%%%%
We have shown that the spin-1 effective Hamiltonians work well for the two-leg and the three-leg Rydberg ladders. For the triangular prism (Sec.~\ref{subsec:triangularprism}) and the triangles in the same plane (Sec.~\ref{subsec:triangleinplane}), the quantum operators are both clock ladder operators $\hat{C}_i^{\pm}$, and $J = \Omega^2V_0/[4\Delta(V_0-\Delta)]$. The effective interaction strengths $D = \Delta_0+2(V_2-V_1), R = (V_1-V_2)/2, R' = 3(V_1-V_2)/2$ for the triangular prism, and $D = \Delta_0 + V_2 + V_4 - 2V_1, R = (V_1-V_3)/2, R' = (3V_1+V_3)/2-V_2-V_4$ for the triangles in the same plane. It is easy to derive the effective Hamiltonians for other geometries and numerically test the validity, we will not shown more results here. The effective Hamiltonian approach can largely reduce the dimension of the Hilbert space for finite-size systems and increase the efficiency of the search for the low-energy quantum states and the study of the dynamical properties of quantum systems numerically on classical computers.

%%%%%%%%%%%%%%%%%%%%%%%%%%%%%%%%%
\section{Phase diagram}
\label{sec:phases}
%%%%%%%%%%%%%%%%%%%%%%%%%%%%%%%%%
\begin{table}[t!]
\caption{\label{tab:orderparams}The values of order parameters for different phases. The corresponding susceptibilities diverge on the critical lines.}
\centering
\begin{tabular} {p{2.cm} p{2.cm} p{2.cm}  p{2.cm}}
    \hline
    \hline
        &$M_{\rm{FM}}$ &$M_{\rm{AFM}}$ &$M_{\rm{RDW}}$
        \\  \hline
    FM &$\neq 0$ &$= 0$ &$= 0$  \\  \hline
    AFM &$= 0$ &$\neq 0$ &$= 0$  \\  \hline
    FRDW &$\neq 0$ &$\neq 0$ &$\neq 0$  \\  \hline
    PRDW &$= 0$ &$= 0$ &$\neq 0$ \\  \hline
    Disorder &$= 0$ &$= 0$ &$= 0$ 
    \\  \hline \hline
    \end{tabular}  
\end{table}
\begin{figure}[t!]
    \centering
    \includegraphics[width=0.5\textwidth]{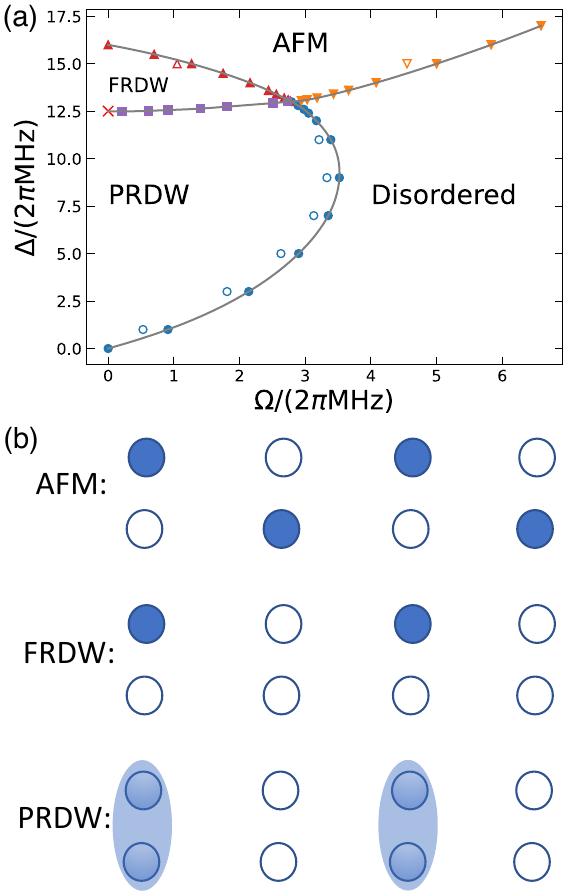}
    \caption{(a) Ground-state phase diagram for the effective Hamiltonian [Eq.~\eqref{eq:effectiveham2legNN}] of the two-leg Rydberg ladder. The phase boundaries between the PRDW and the disordered phases, between the PRDW and the FRDW phases, between the FRDW and the AFM phases, and between the AFM and the disordered phases are determined by the peak positions of susceptibilities $\chi_{\rm{R}}$, $\chi_{\rm{F}}$, $\chi_{\rm{A}}$, and $\chi_{\rm{A}}$ with $N_s = 512$, respectively. Here, $V_0 = 1000\times 2\pi$ MHz and $\rho = a_y/a_x= 0.5$. The empty markers are the corresponding peak positions of susceptibilities in the real two-leg Rydberg ladder. The red cross at $\Omega = 0$ is the analytical value of the phase transition point between the PRDW and the FRDW phases in the $\Omega \rightarrow 0$ limit. (b) The density maps for the three ordered phases. The filled and the empty circles represent the Rydberg excited states and the single-atom ground states, respectively. The shaded rungs in the PRDW phase represent entangled bonds without up-down symmetry breaking.}
    \label{fig:gsphasediag2leg}
\end{figure}

We take the two-leg Rydberg ladder as an example to study the ground-state phase diagram of the Rydberg ladder system. In the effective Hamiltonian~\eqref{eq:effectivehamgeneral}, when $D$ and $R'$ are large, all spins are zero, which is the trivial large-$D$ phase \cite{PhysRevB.67.104401}. For large $J$, the system is in the disordered phase. When $R$ is large, the anti-ferromagnetic (AFM) ordered phase with $\langle \hat{L}_i^z\rangle = -\langle \hat{L}_{i+1}^z\rangle$ is favored. When $R$ is negative and large in magnitude, the ferromagnetic (FM) ordered phase with all $\langle \hat{L}_i^z\rangle$ taking the same value is favored. When $R'$ is large and $D$ is negative, a new Rydberg density wave (RDW) order may emerge: the values of $\langle (\hat{L}^z_i)^2 (\hat{L}^z_{i+1})^2 \rangle$ is minimized and the values of $\langle (\hat{L}^z_i)^2 \rangle$ in the system take the form $\ldots, f, t, f, t, f, t, \ldots$ ($0 \le t < f \le 1$). The order parameters are 
\begin{eqnarray}
\nonumber M_{\rm{FM}} &=& \frac{1}{N_s} \sum_i \langle\hat{L}^z_i\rangle, \\ \nonumber M_{\rm{AFM}} &=& \frac{1}{N_s} \sum_i \left(-1\right)^{i} \langle\hat{L}^z_i\rangle, \\ M_{\rm{RDW}} &=& \frac{1}{N_s} \sum_i \left(-1\right)^i \langle (\hat{L}^z_i)^2 \rangle
\end{eqnarray}
for FM, AFM, and RDW, respectively. The susceptibilities for the three order parameters are defined by 
\begin{eqnarray}
\nonumber \chi_{\rm{F}} &=& \frac{1}{N_s} \sum_{i,j}\left(\langle \hat{L}_i^z \hat{L}_j^z \rangle - \langle \hat{L}_i^z \rangle \langle \hat{L}_j^z \rangle\right), \\ \nonumber \chi_{\rm{A}} &=& \frac{1}{N_s} \sum_{i,j}\left(-1\right)^{i+j}\left(\langle \hat{L}_i^z \hat{L}_j^z \rangle - \langle \hat{L}_i^z \rangle \langle \hat{L}_j^z \rangle\right), \\ \nonumber \chi_{\rm{R}} &=& \frac{1}{N_s} \sum_{i,j}\left(-1\right)^{i+j}\left[\langle (\hat{L}_i^z)^2 (\hat{L}_j^z)^2 \rangle - \langle (\hat{L}_i^z)^2 \rangle \langle (\hat{L}_j^z)^2 \rangle\right], \\
\end{eqnarray}
respectively. At the phase transition point between the ordered phase and the disordered phase, the associated susceptibility diverges if the phase transition is of second order. Bearing in mind that the spin operators and the Rydberg state operators are related by Eqs. \eqref{eq:number} and \eqref{eq:oprelations1} is helpful in the following analysis.
\begin{figure}[t!]
    \centering
    \includegraphics[width=0.5\textwidth]{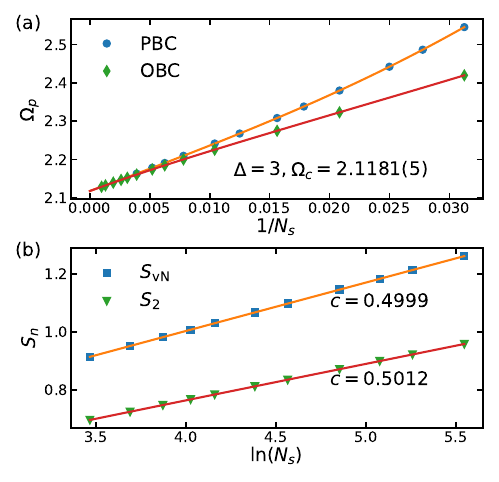}
    \caption{(a) Finite-size scaling of the peak position of the susceptibility $\chi_{\rm{R}}$ of the ground state on the $\Delta = 3\times 2\pi$ MHz cut for the effective Hamiltonian~\eqref{eq:effectiveham2legNN} of the two-leg Rydberg ladder. Changing $\Omega$ drives the phase transition between the PRDW and the disordered phases. Both PBCs and OBCs are considered. Almost the same extrapolated values of the critical point for the two boundary conditions are obtained and $\Omega_c = 2.1181(5)\times 2\pi$ MHz. The Hamiltonian parameters in the plot are in units of $2\pi$ MHz. (b) The von Neumann entanglement entropy $S_{\rm{vN}}$ and the second-order R\'enyi entropy $S_2$ for PBCs as functions of $\ln(L)$ are plotted at $\Omega_c=2.1181$. The extracted values of the central charge $c$ are close to $0.5$ predicted by Ising CFT.}
    \label{fig:fssdel3pcdwtodis2leg}
\end{figure}

The possible phases with different values of the order parameters are summerized in Table~\ref{tab:orderparams}. In the FM phase, only $M_{\rm{FM}}$ is nonzero and the up-down $\mathbb{Z}_2$ symmetry is broken. Similarly, only $M_{\rm{AFM}}$ is nonzero in the AFM phase. In the RDW phase, $M_{\rm{RDW}}$ is nonzero, and if the other two order parameters are both zero, we have the paramagnetic RDW (PRDW) phase. In the PRDW phase, $\langle \hat{L}_i^z\rangle = 0$ for all the sites but $\langle (\hat{L}_i^z)^2\rangle \neq 0$ for even or odd sites, which means all the rungs in the Rydberg ladder preserve the up-down $\mathbb{Z}_2$ symmetry and only the translational $\mathbb{Z}_2$ symmetry is broken. Because the states with two Rydberg states in a rung is blockaded, the single-rung state in the PRDW phase should have large overlap with the entangled state $\left(\ket{n_{i,+1}=1,n_{i,-1}=0} - \ket{n_{i,+1}=0,n_{i,-1}=1}\right)/\sqrt{2}$. Similar phases with a translational $\mathbb{Z}_3$ symmetry breaking are also found in a two-leg square Rydberg ladder \cite{EckFendleyPRB2023}. We remark that the state with AFM order in even or odd sites also satisfies the above condition for the PRDW phase, but it requires strong NNN interactions that do not exist in our effective Hamiltonian. If all the three order parameters are nonzero, we have the ferromagnetic RDW (FRDW) phase where a FM phase exists in the sublattice with even or odd sites and other sites are empty. The schematic views of the AFM, the FRDW, and the PRDW phases are shown in Fig.~\ref{fig:gsphasediag2leg}(b).

We calculate the order parameters and the associated susceptibilities for the effective Hamiltonian~\eqref{eq:effectiveham2legNN} with $N_s=512$, $V_0 = 1000\times 2\pi$ MHz, and $\rho = 0.5$. As shown in Fig.~\ref{fig:gsphasediag2leg}(a), three ordered phases, the AFM, the FRDW, and the PRDW phases are identified in the $\Delta$-$\Omega$ plane. The peak positions of the susceptibilities are used to estimate the phase transition lines. In practice, we use the peak of $\chi_{\rm{R}}$ to estimate the phase transition line between the PRDW and the disordered phases and the phase transition line between the FRDW and the AFM phases. Notice that from the FRDW phase to the AFM phase, the lattice translational symmetry is restored from $\mathbb{Z}_2$ symmetry breaking for $\langle (\hat{L}_i^z)^2\rangle$, so the susceptibility $\chi_{\rm{R}}$ can also detect the phase transition point. From the FRDW phase to the PRDW phase, the FM order on even or odd sites vanishes, so we use $\chi_{\rm{F}}$ to locate the phase boundary. Finally, we use $\chi_{\rm{A}}$ to locate the phase boundary between the AFM and the disordered phases. The empty markers in Fig.~\ref{fig:gsphasediag2leg}(a) are the results for the real two-leg Rydberg ladder $\hat{H}_{\rm{2LR}}$, which are consistent with those for the effective Hamiltonian.

\begin{figure}[t!]
    \centering
    \includegraphics[width=0.5\textwidth]{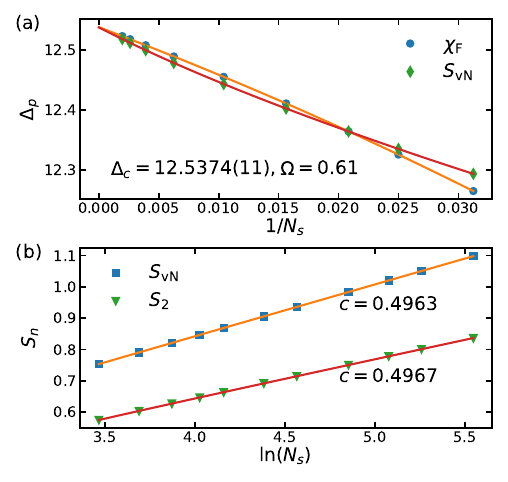}
    \caption{(a) Finite-size scaling of the peak position of the susceptibility $\chi_{\rm{A}}$ and that of the von Neumann entanglement entropy $S_{\rm{vN}}$ for the ground state of Hamiltonian~\eqref{eq:effectiveham2legNN} with OBCs on the $\Omega = 0.61\times 2\pi$ MHz cut. Varying $\Delta$ drives the phase transition between the PRDW and the FRDW phases. The extrapolated values of the critical point from $\chi_{\rm{A}}$ and $S_{\rm{vN}}$ are almost the same and $\Delta_c = 12.5374(11)\times 2\pi$ MHz is obtained. The Hamiltonian parameters in the plot are in units of $2\pi$ MHz. (b) Same as Fig.~\ref{fig:fssdel3pcdwtodis2leg}(b), but for $\Delta_c = 12.5374(11)\times 2\pi$ MHz and $\Omega = 0.61\times 2\pi$ MHz.}
    \label{fig:fssomg0p61pcdwtofcdw2leg}
\end{figure}
\begin{figure}[t!]
    \centering
    \includegraphics[width=0.49\textwidth]{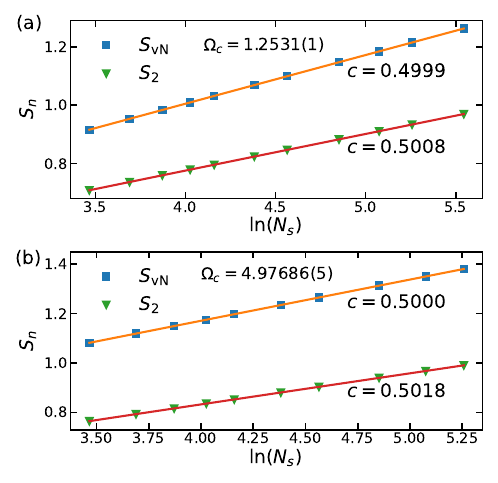}
    \caption{Same as Fig. \ref{fig:fssdel3pcdwtodis2leg}(b) and Fig. \ref{fig:fssomg0p61pcdwtofcdw2leg}(b), but for (a) the critical point between the FRDW and the AFM phases [$\Delta = 15, \Omega_c = 1.2531(1)$] and (b) the one between the AFM and the disordered phases [$\Delta = 15, \Omega_c = 4.97686(5)$].}
    \label{fig:fssdel15fcdwtofm2leg}
\end{figure}
The existence of the ordered phases in each parameter regime can be understood from the classical limit. When $\Omega = 0$, the Hamiltonian~\ref{eq:effectiveham2legNN} is diagonal. By comparing the energy density of the possible orders, one knows that the system is in the large-$D$ phase ($\ket{\ldots,0,0,0,0,\ldots}$) for $\Delta < 0$, in the RDW phase ($\ket{\ldots, \pm 1, 0, \pm 1, 0, \ldots}$) for $0 < \Delta < 2V_2$, and in the AFM phase ($\ket{\ldots, +1, -1, +1, -1, \ldots}$) for $\Delta > 2V_2$. Because $0< R < R'$ independent of $\Delta$ or $\Omega$, the $(\hat{L}_i^z)^2(\hat{L}_{i+1}^z)^2$ interaction term dominates and even or odd sites should be empty to minimize the interaction energy. Then the large-$D$ phase is favored for positive $D$ and the RDW phase is favored for negative $D$ with small magnitude. When $D = -\Delta$ is negative with large magnitude, $(L_i^z)^2 = 1$ is preferred for each site, and the positive $R$ picks out the AFM phase. The RDW phase at $\Omega = 0$ has exponentially large degeneracy. When $\Omega$ is turned on a little, an equal superposition of $\ket{+1}$ and $\ket{-1}$ is favored in the presence of the operator $\hat{U}^+_i + \hat{U}^-_i$, so the PRDW phase shows up. In the presence of quantum fluctuations ($\Omega > 0$), the NNN nonzero spin states can also interact via the middle spin in a perturbative way. One can perform degenerate perturbation theory for small $\Omega$ and obtain an effective Ising model on even or odd sites:
\begin{equation}
\label{eq:effIsinginRDW}
\hat{H}^{\rm{eff}}_{\rm{Ising}} = \frac{\Omega^2}{4} \left(-J_{\rm{eff}}\sum_{i} \sigma^z_{i}\sigma^z_{i+2} - \frac{1}{\Delta} \sum_i \sigma^x_i\right). 
\end{equation}
with 
\begin{equation}
J_{\rm{eff}}=\frac{1}{\Delta-V_1-V_2} - \frac{1}{2\Delta-4V_2} - \frac{1}{2\Delta-4V_1}
\end{equation}
We notice that $J_{\rm{eff}}$ is always positive for $0 < \Delta < 2V_2$. So the PRDW phase can spontaneously break the up-down $\mathbb{Z}_2$ symmetry to have a FM order in even or odd sites. In this case, we have the FRDW phase. In a word, the nonzero quantum fluctuations remove the macroscopically large degeneracy of the classical ground states and stabilize a definite order (the PRDW or the FRDW orders here), which is called the order-by-disorder mechanism \cite{villain1980order}. This phenomenon is also observed in a two-leg Rydberg ladder with staggered detuning \cite{SarkarOrderDisorder2023}. The Ising critical point can be determined by $J_{\rm{eff}} = 1/\Delta$, which gives $\Delta_c \approx 12.509 \times 2\pi$ MHz. This analytical Ising critical point in the classical limit is highlighted by the red cross in the phase diagram in Fig.~\ref{fig:gsphasediag2leg}(a). We extrapolate the numerical phase transition points between the FRDW and the PRDW phases to the value at the $\Omega = 0$ limit, which agrees perfectly with the analytical result.

Across each phase transition line, only one $\mathbb{Z}_2$ symmetry is spontaneously broken, we expect that all the phase transitions belongs to the Ising universality class. We can use the entanglement entropy to detect the Ising phase transition. If the ground state is $\ket{\Psi}$, the first-order R\'enyi entropy or the von Neumann entropy $S_{\rm{vN}} = - \Tr \rho_{\mathcal{A}} \ln \rho_{\mathcal{A}}$, where $\rho_{\mathcal{A}} = \Tr_{\mathcal{B}} \ket{\Psi}\bra{\Psi}$ is the reduced density matrix for the subsystem $\mathcal{A}$ if the system is partitioned into $\mathcal{A}$ and $\mathcal{B}$ parts. The $n$-th order R\'enyi entropy can be calculated by $S_n = 1/\left(1-n\right)\ln \Tr \rho^n_{\mathcal{A}}$. We consider the cut is at the middle and subsystems $\mathcal{A}$ and $\mathcal{B}$ have the same volume. The conformal field theory (CFT) predicts that the central charge $c = 0.5$ for the Ising phase transition \cite{BELAVIN1984333}, and the $n$th-order R\'enyi entanglement entropy at the critical point diverges logarithmically with a coefficient related to the central charge \cite{AffleckGSDeg1991,PhysRevLett.90.227902,Calabrese_2004,HOLZHEY1994443,Calabrese_2009}
\begin{eqnarray}
\label{eq:ent}
S_{n}= \begin{cases} \frac{c\left(n+1\right)}{12n} \ln \left(N_s\right)+s_{o} & \textrm{for OBC} \\ \frac{c\left(n+1\right)}{6n} \ln \left(N_s\right)+s_{p} & \textrm{for PBC},
\end{cases}
\end{eqnarray}
where $s_{o(p)}$ is a nonuniversal constant, OBC and PBC stand for open boundary condition and periodic boundary condition, respectively.

\begin{figure}[t!]
\includegraphics[width=8.6cm]{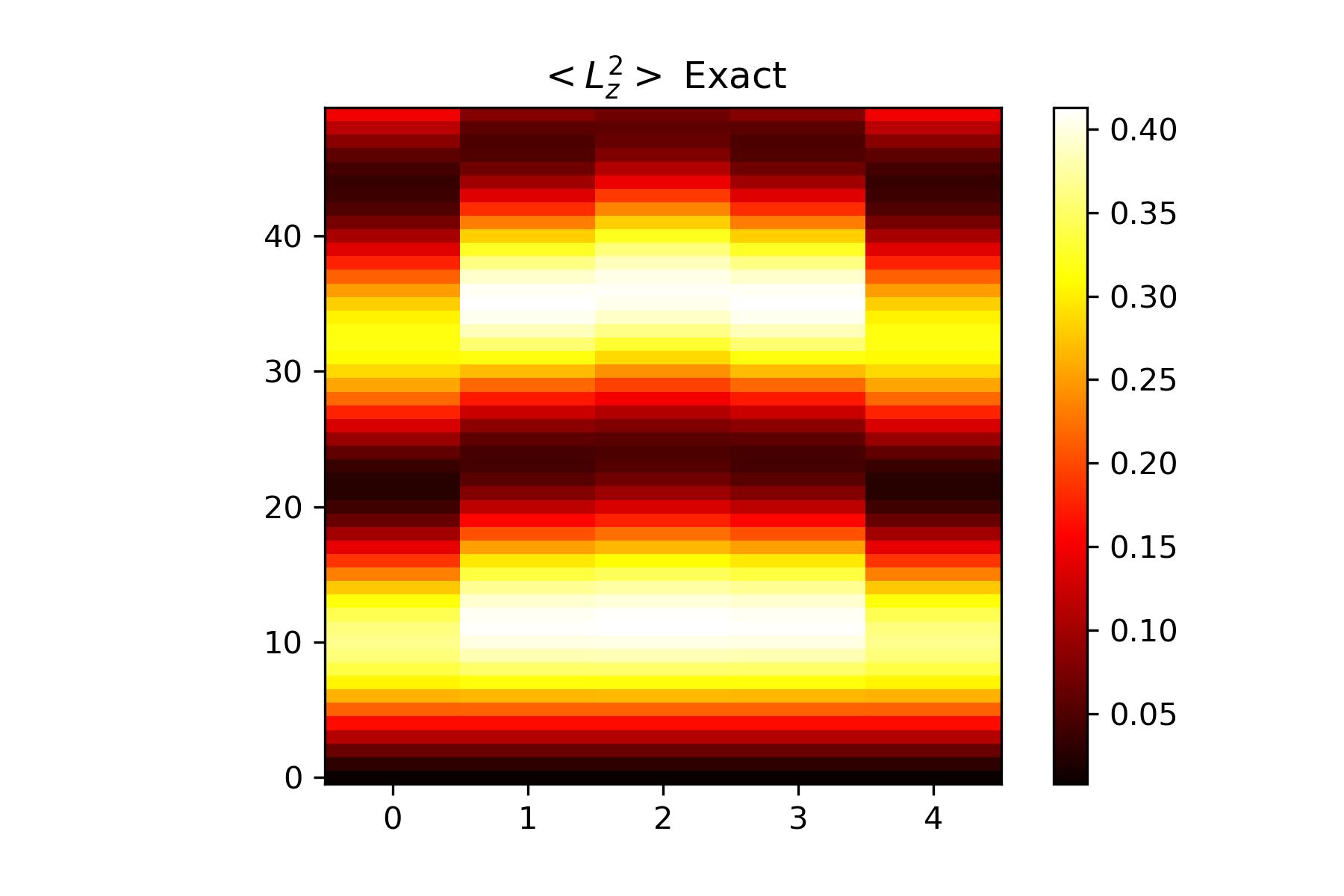}

\includegraphics[width=6.cm]{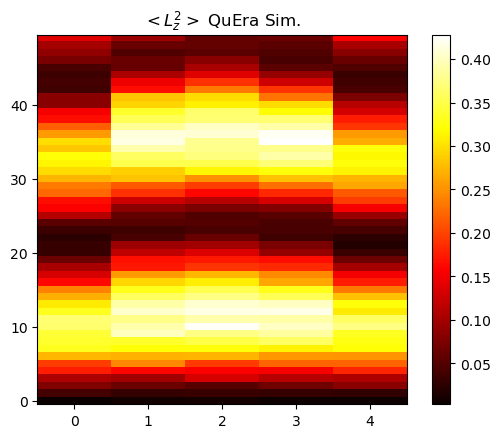}

\includegraphics[width=8.6cm]{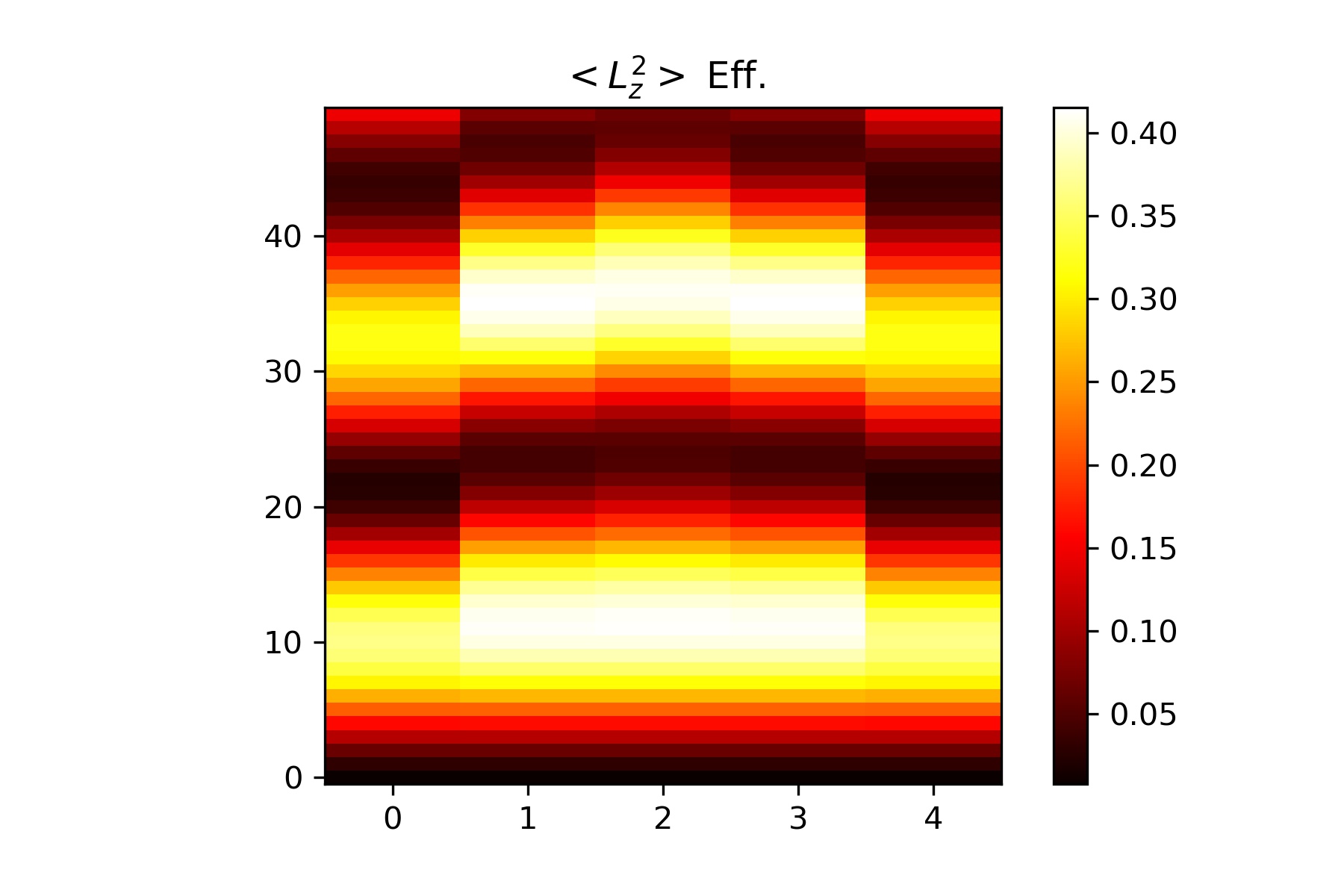}
\caption{\label{fig:5sitesV} Values of $\langle(\hat{L}^z_i)^2\rangle$ for five sites, $\Omega=4\pi$ MHz, $\Delta=2\Omega$, $\rho=0.5$ $a_x=1R_b$ and $a_y=0.5\times R_b$. The vertical time units are $10^{-8}$ s. Top:exact diagonalization. Middle: QuEra (local simulator). Bottom: Effective Hamiltonian.}
\end{figure}
\begin{figure}[t!]
\includegraphics[width=8cm]{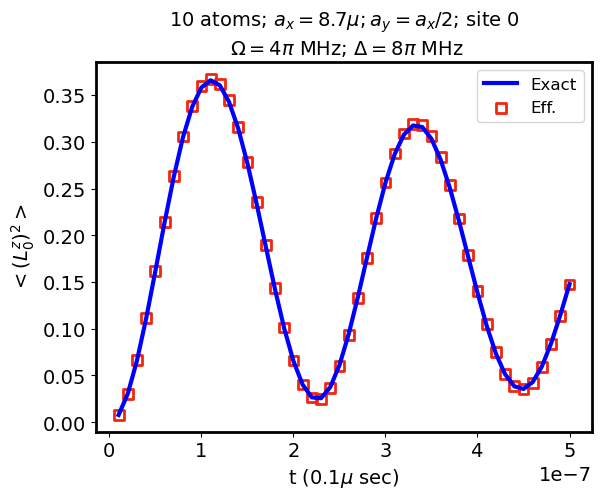}
\includegraphics[width=8cm]{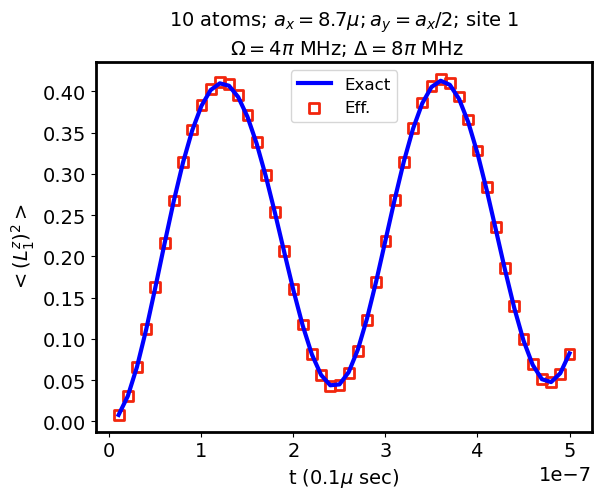}
\includegraphics[width=8cm]{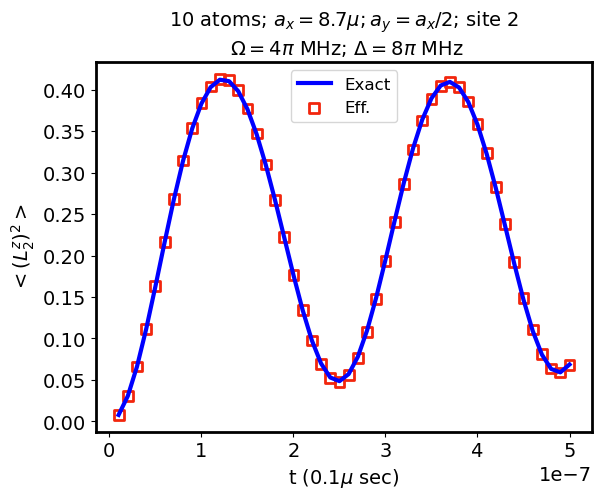}
\caption{\label{fig:5sitesE} Values of $\langle(\hat{L}^z_i)^2\rangle$ for the first three of the five sites. The exact and effective evolution shown in Fig. \ref{fig:5sitesV} are compared here.}
\end{figure}
\begin{figure}[t!]
\includegraphics[width=8.6cm]{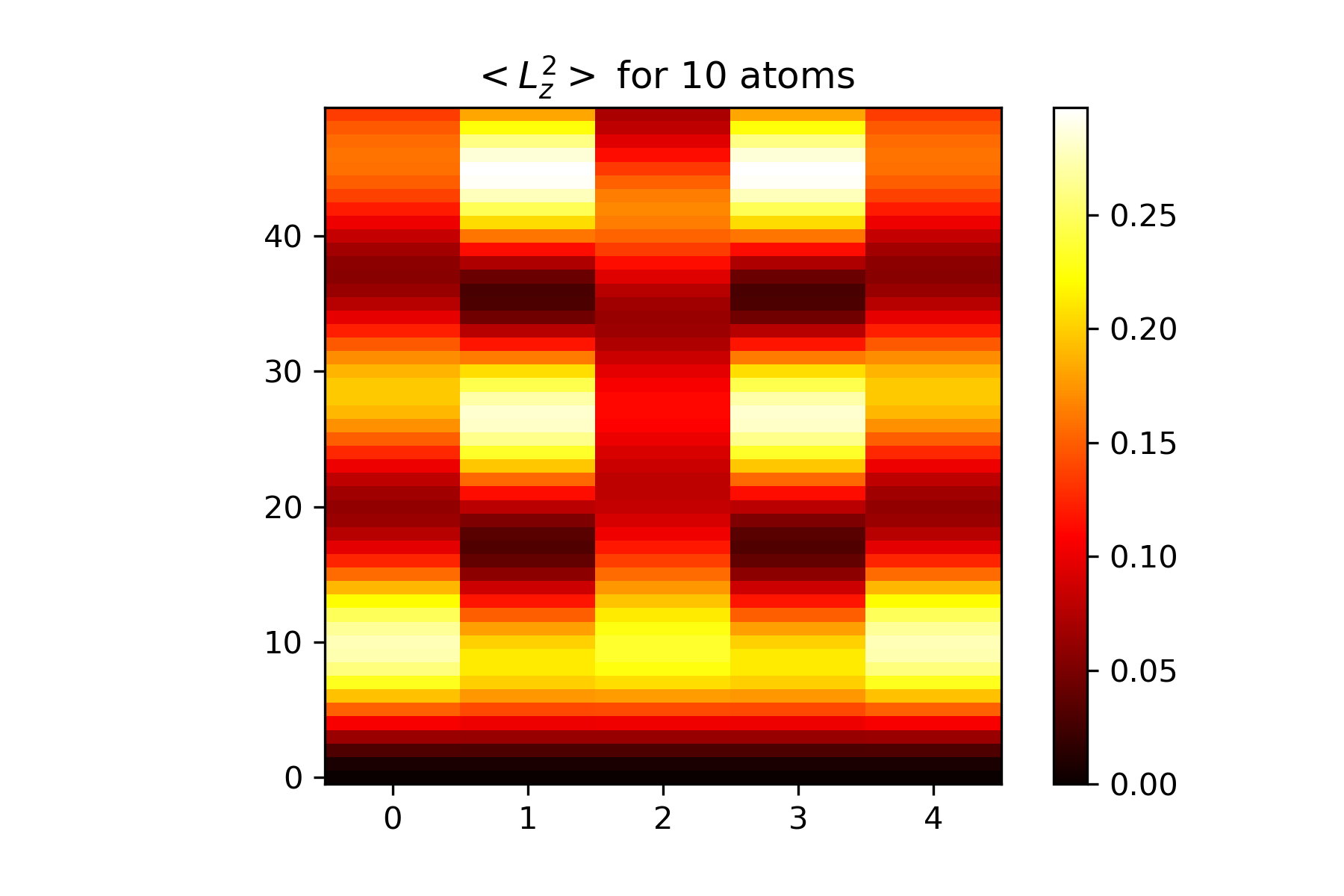}
\includegraphics[width=8.6cm]{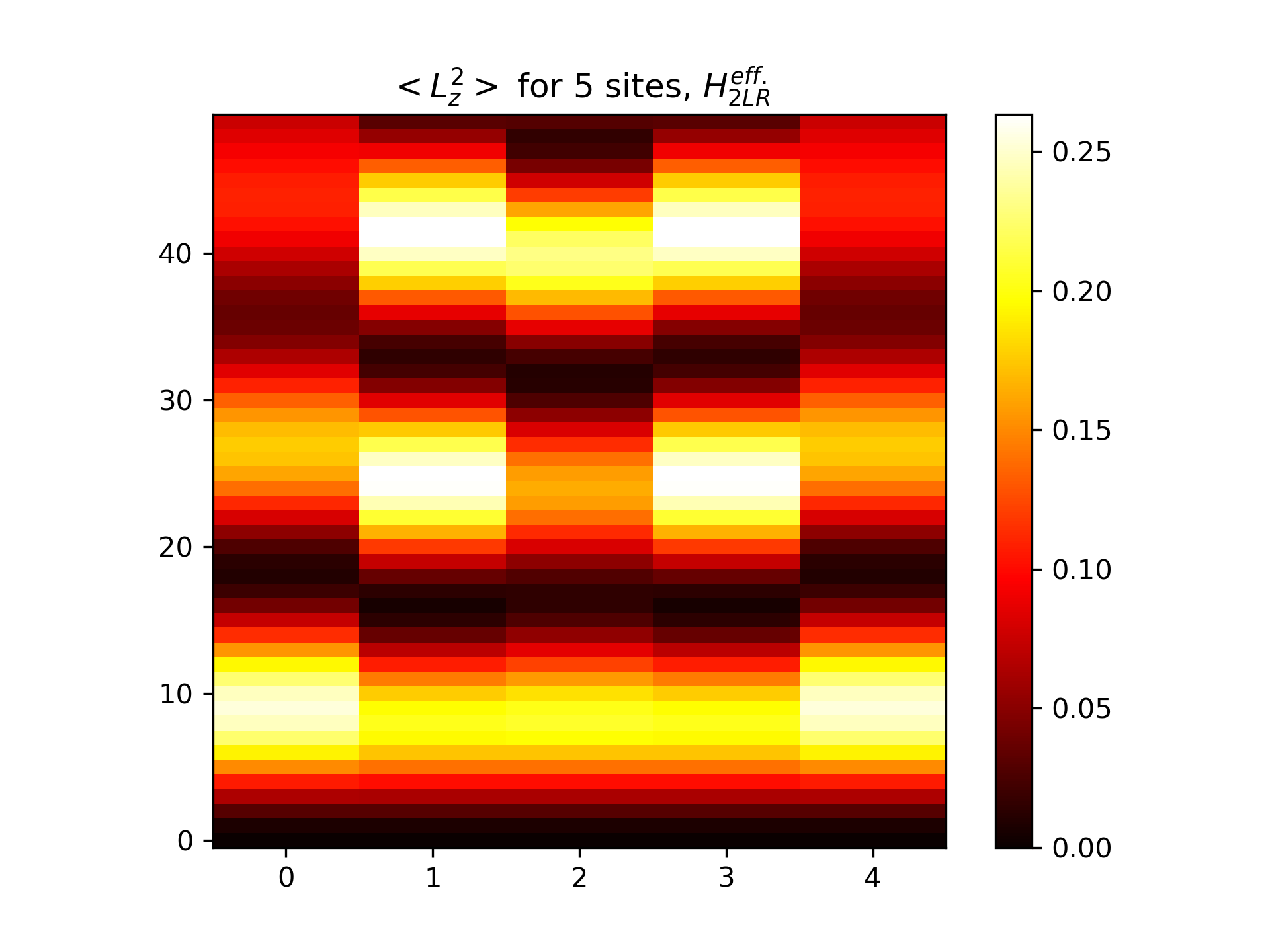}
\caption{\label{fig:out1V}5 sites, $\rho=0.5$, $a_x=0.5\times R_b$, $\Omega=4\pi$ MHz, $\Delta=2\Omega$, vertical time units are $0.01\mu$sec. Top: 10 atom simulator. Bottom: 5 site $\hat{H}^{\rm{eff}}_{\rm{2LR}}$.}
\end{figure}
\begin{figure}[t!]
\includegraphics[width=8.6cm]{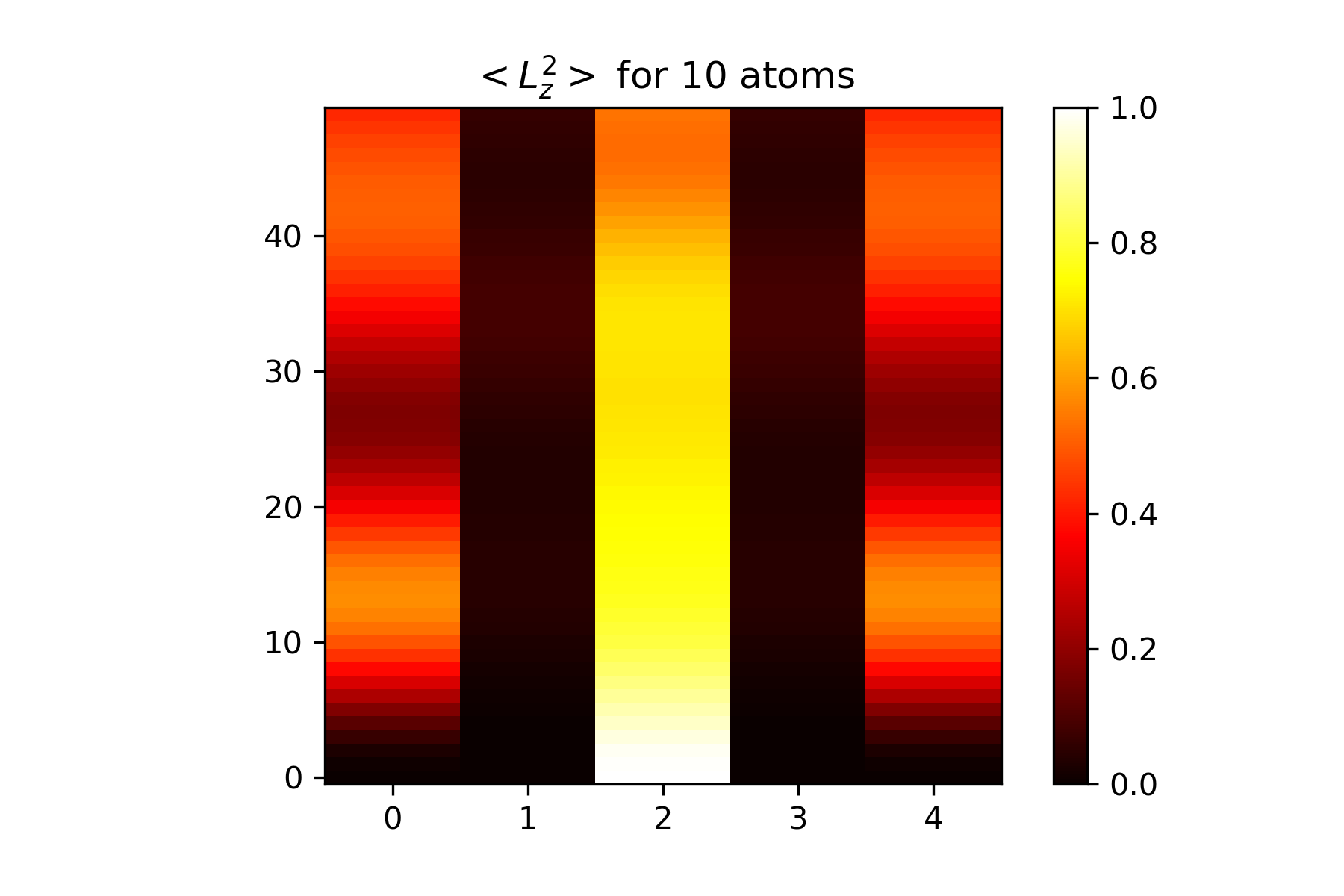}
\includegraphics[width=8.6cm]{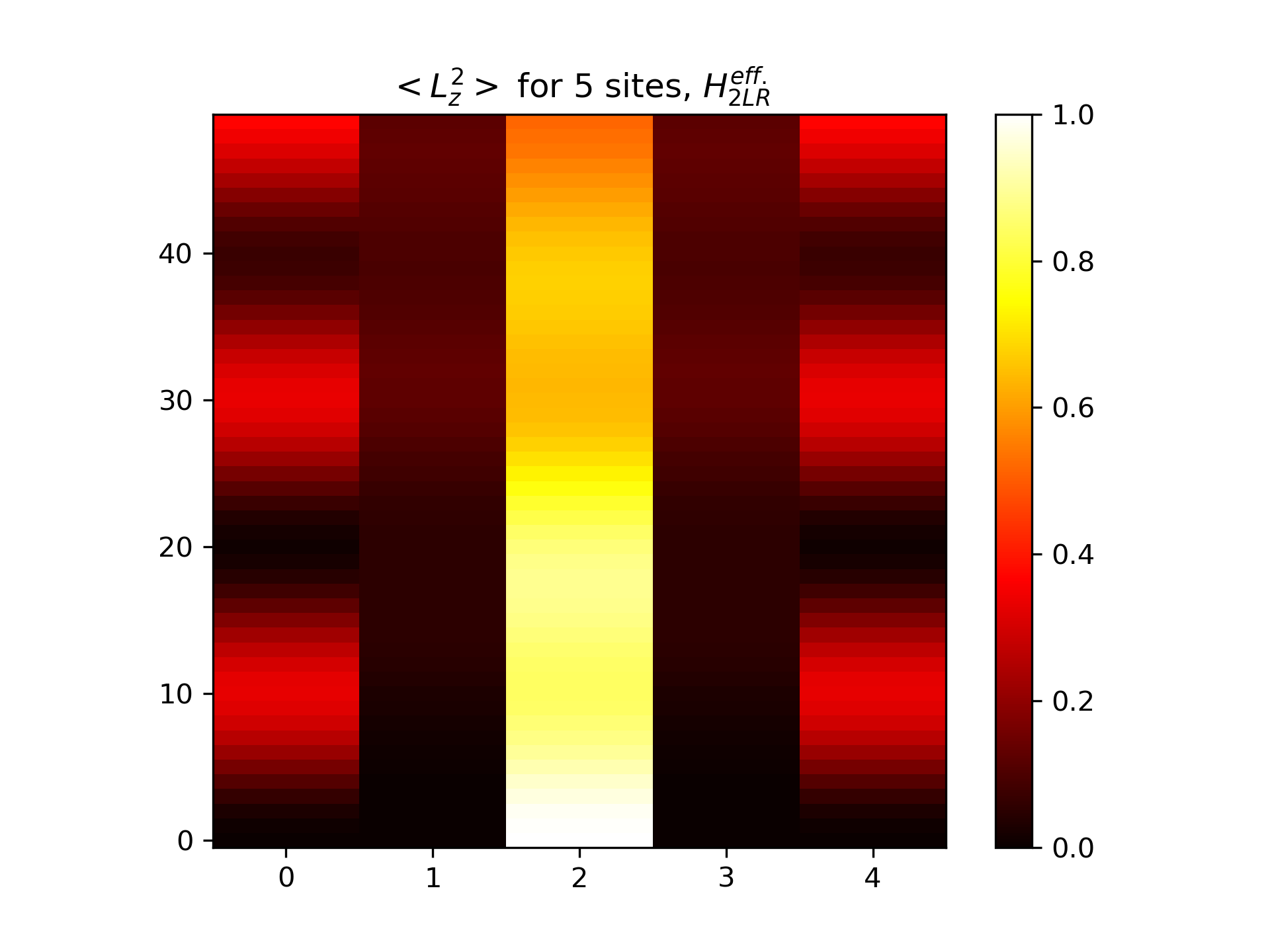}
\caption{\label{fig:out1E}5 sites, $\rho=0.5$, $a_x=0.5\times R_b$, $\Omega=4\pi$ MHz, $\Delta=2\Omega$, vertical time units are $0.01\mu$sec. Top: 10 atom simulator. Bottom: 5 site $\hat{H}^{\rm{eff}}_{\rm{2LR}}$.}
\end{figure}
\begin{figure}[t!]
\includegraphics[width=8.6cm]{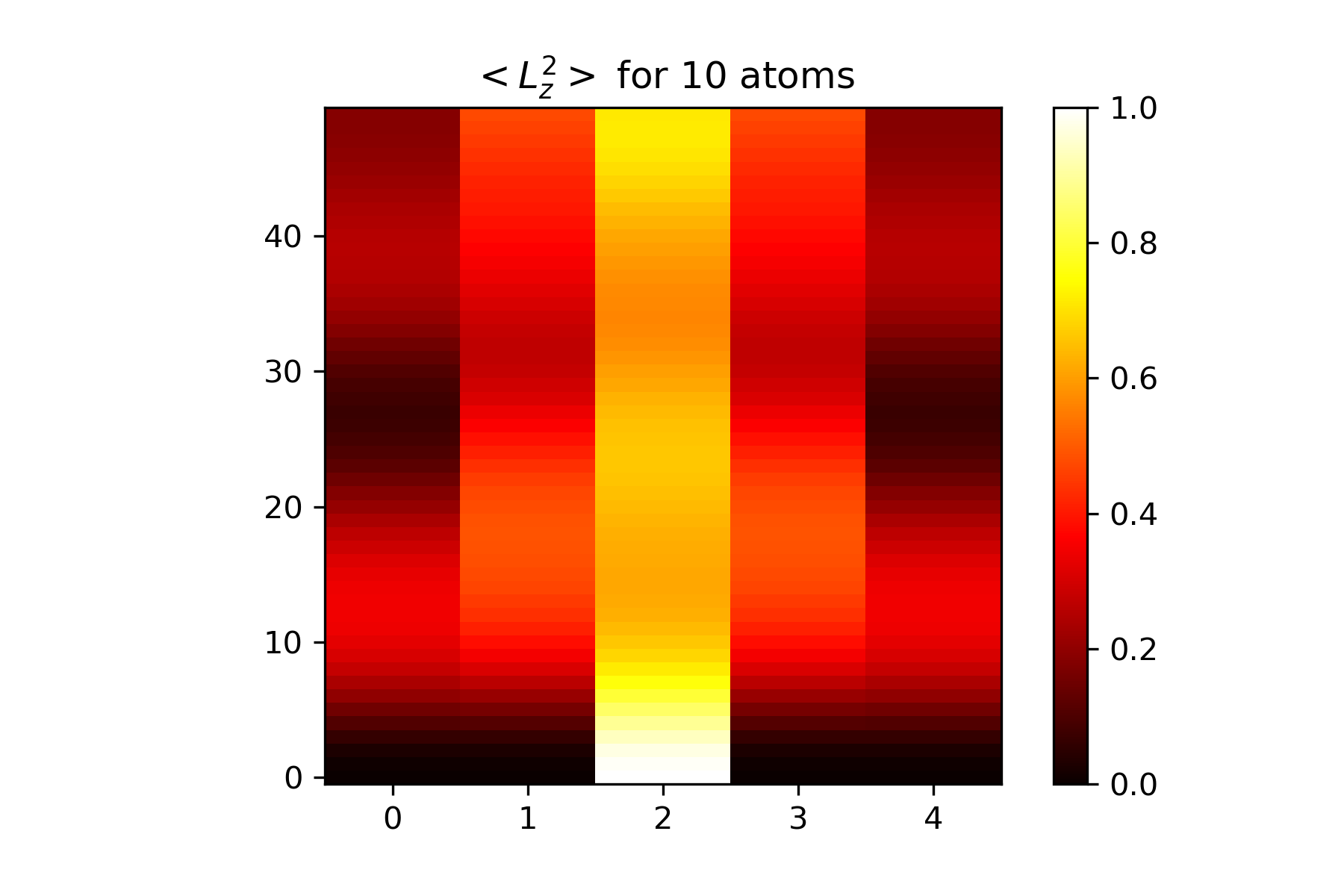}
\includegraphics[width=8.6cm]{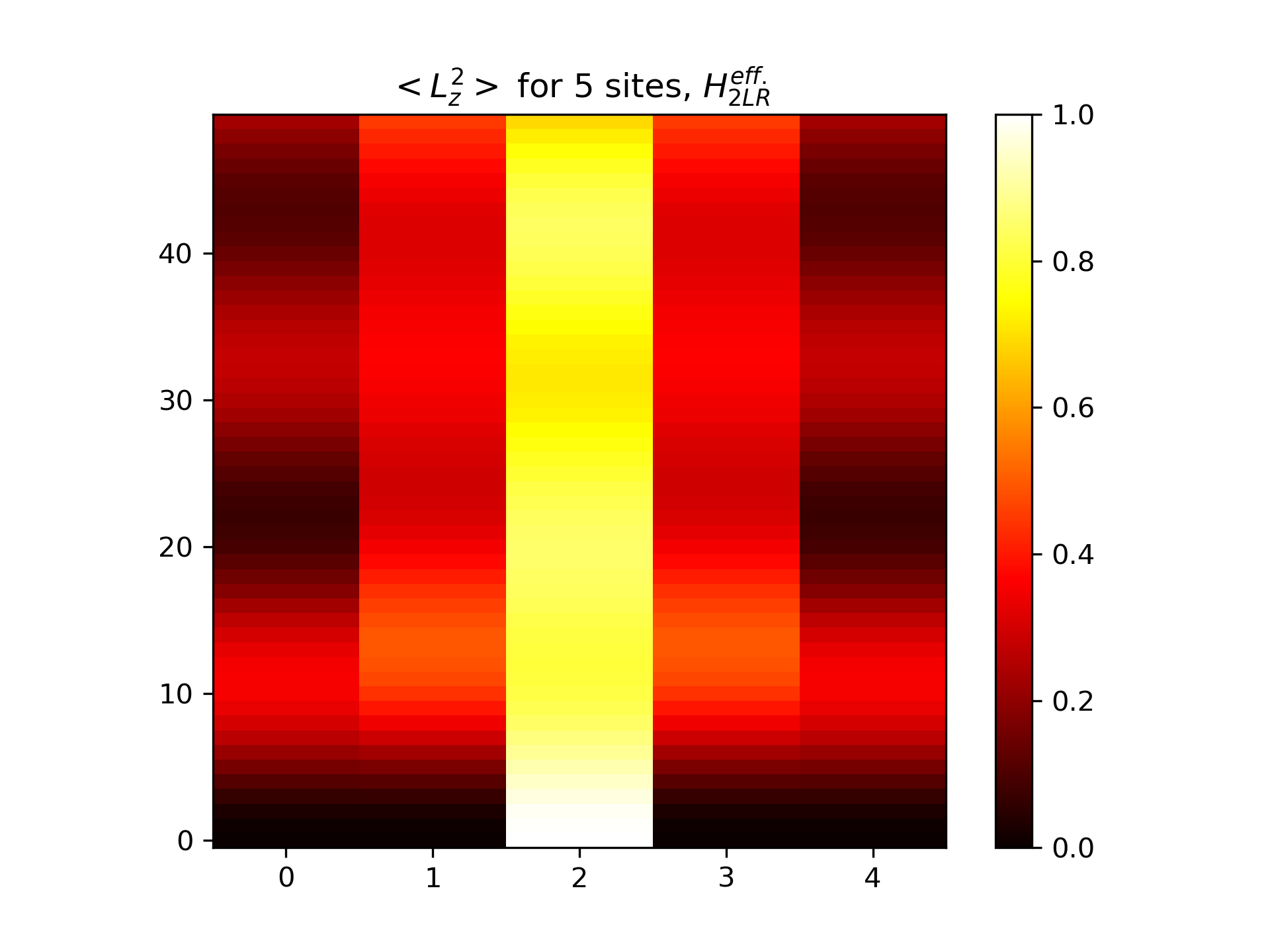}
\caption{\label{fig:out2E}5 sites, $\rho=2$, $a_x=1\times R_b$, $\Omega=4\pi$ MHz, $\Delta=2\Omega$, vertical time units are $0.01\mu$sec. Top: 10 atom simulator. Bottom: 5 site $\hat{H}^{\rm{eff}}_{\rm{2LR}}$.}
\end{figure}

In order to calculate the central charge, it is required to obtain the critical point in the thermodynamic limit first. For the phase transition point between the PRDW and the disordered phases, we consider the $\Delta = 3\times 2\pi$ MHz cut and calculate the susceptibility $\chi_{\rm{R}}$ in a small window around the peak and find the peak position $\Omega_p$ for different system sizes. The results for both PBC and OBC are presented in Fig.~\ref{fig:fssdel3pcdwtodis2leg}(a). The data for PBC and OBC becomes indistinguishable for large $N_s$. We use a polynomial to fit $\Omega_p$ as a function of $1/N_s$ and extrapolate the value of $\Omega_p$ to the $N_s\rightarrow \infty$ limit. Then we obtain the critical point $\Omega_c = 2.1181(5)\times 2\pi$ MHz, where the uncertainty is estimated from the error of curve fit and the discrepancy between results of PBC and OBC. In Fig.~\ref{fig:fssdel3pcdwtodis2leg}(b), we show the von Neumann entropy $S_{\rm{vN}}$ and the second-order R\'enyi entropy as functions of $\ln(N_s)$ for PBCs at $\Omega_c$. The entropy data is fit to the CFT form in Eq.~\eqref{eq:ent} and the extracted values of the central charge are $c = 0.4999$ and $c = 0.5012$ for $S_{\rm{vN}}$ and $S_2$, respectively. Notice that we add a $1/N_s^2$ correction term in the curve fit to increase the accuracy \cite{PhysRevD.96.034514}. The results perfectly agree with the expected value $c=0.5$ for Ising CFT. We also fit the peak height of $\chi_{\rm{R}}$ to the power-law form $aN_s^{1-\eta}+b$ and obtain $\eta = 0.257$ and $0.248$ for OBCs and PBCs, respectively, which are consistent with $\eta=1/4$ for the Ising universality class.

We next consider the $\Omega = 0.61\times 2\pi$ MHz cut and study the phase transition between the PRDW and the FRDW phases in Fig.~\ref{fig:fssomg0p61pcdwtofcdw2leg}. In this case, the susceptibility $\chi_{\rm{F}}$ and the von Neamann entropy $S_{\rm{vN}}$ for OBCs are calculated and the finite size scalings of the peak positions $\Delta_p$ are used for the extrapolation of the critical point. We also obtain almost the same value of $\Delta_c$ from $\chi_{\rm{F}}$ and $S_{\rm{vN}}$ and $\Delta_c = 12.5374(11)\times 2\pi$ MHz. The numerical values of the central charge $c = 0.4963$ and $c = 0.4967$ for $S_{\rm{vN}}$ and $S_2$, respectively. We perform the same procedure for the $\Delta = 15\times 2\pi$ MHz cut, where there are two critical points $\Omega_c = 1.2531(1)\times 2\pi$ MHz between the FRDW and the AFM phases, and $\Omega_c = 4.97686(5)\times 2\pi$ MHz btween the AFM and the disordered phases. The results for the entanglement entropy at the critical points are presented in Fig.~\ref{fig:fssdel15fcdwtofm2leg}. One can see that the numerical values of the central charge are all consistent with the Ising CFT prediction $c=0.5$. We also obtain $\eta=0.249, 0.251$, and $0.252$ for the three critical points, respectively, which are all cosistent with $\eta=1/4$ for Ising university class.

In summary, the effective spin-1 Hamiltonian in Eq.~\eqref{eq:effectiveham2legNN} for the two-leg Rydberg ladder with the inverse aspect ratio $\rho=0.5$ has a rich phase diagram with three density wave orders and a disordered phase. The phase boundaries between these phases are all Ising critical lines. We remark that both the PRDW phase and the AFM phase spontaneously break one translational $\mathbb{Z}_2$ symmetry and the ground states have two-fold degeneracy, while the FRDW phase breaks two different $\mathbb{Z}_2$ symmetries and the ground state has four-fold degeneracy. The phase transition from the FRDW phase to the disordered phase can only take place across the intersection point of the two Ising critical lines, which is in the universality class of the four-state clock model or two uncoupled Ising models. Our results suggest to study $\Delta >0$ as an environment effect, staggered structures (cells with two sites) and to explore outside the region of validity of $\hat{H}^{\rm{eff}}_{\rm{mLR}}$.

%%%%%%%%%%%%%%%%%%%%%%%%%%%%%%%%%
\section{Practical Applications}
\label{sec:practical}
%%%%%%%%%%%%%%%%%%%%%%%%%%%%%%%%%%%%

In this section, we discuss simple examples of real-time evolution for two leg ladders with 10 atoms (5 sites). This situation can be implemented remotely using QuEra facilities or simulated with a local SDK \cite{quera,aws}. In this case, the system is initialized with the 10 atoms in the ground state. This initial state is invariant under the transformation $\hat{L}^z_i \rightarrow -\hat{L}^z_i$ or equivalently swapping the two legs of the ladder and as we turn on the evolution the $\langle\hat{L}^z_i\rangle$ remain zero. For this reason, we display the values of $\langle(\hat{L}^z _i)^2\rangle$ for the five sites which is also invariant under staggered redefinitions of $\hat{L}_i^z$. Note also that the results are left-right symmetric. Results for the typical QuEra values 
$\Omega=4\pi$ MHz, $\Delta=2\Omega$, $\rho=0.5$ and the distance between the sites $a_x=1R_b\simeq 8.7 \mu m$ are shown below. The distance between the two atoms on one site is $a_y=0.5 R_b$ so $\ket{rr}$ at that site are unlikely. The evolution for a period 0.5 $\mu$s is shown in Fig. \ref{fig:5sitesV}, with exact diagonalization for the 10 site problem, the QuEra local simulator and the effective Hamiltonian. It appears clearly that the three methods give very similar results. A finer comparison for the first three sites in Fig. \ref{fig:5sitesE} confirms that the effective Hamiltonian results are in excellent agreement. 

We have also explored values of the parameters that are slightly outside the range of validity of $\hat{H}^{\rm{eff}}_{\rm{2LR}}$. As a first example we keep $\rho=0.5$, but we reduce  the lattice spacing to $a_x=R_b/2$. The blockade mechanism is even more effective but $a_x=R_b/2$ implies next to nearest neighbor interactions not taken into account in the simple $\hat{H}^{\rm{eff}}_{\rm{2LR}}$ with just NN interactions as in Eq. (\ref{eq:effectiveham2legNN}). Figure~\ref{fig:out1V} makes it clear that the effective description is a less accurate description of the simulator than in the previous case, however, there is a qualitative agreement between the exact diagonalization with 10 atoms and the corresponding $\hat{H}^{\rm{eff}}_{\rm{2LR}}$. It should also be noticed that for $a_x=R_b/2$, $V_1$ and $V_2$ are 64 times larger than those in the previous case and the quartic term becomes very important. This is signaled by $L_z=0$ bands screening the electric field both in the effective theory and the original simulator.
This is illustrated in Fig. \ref{fig:out1E}.

The screening effect observed in the previous deformation can be partially remediated  by rather increasing $\rho$ in order to have $V_2<<V_1$.
For this reason we considered the case $\rho=2$, $a_x=R_b$. The results are shown in Fig. \ref{fig:out2E}. In the original model, the Rydberg blockade radius $R_b$ is smaller than the length of rungs, so a Rydberg state at rung 2 will not forbid nearby rungs having Rydberg states.

%%%%%%%%%%%%%%%%%%%%%%%%%%%%%%%%%
\section{Conclusion}
%%%%%%%%%%%%%%%%%%%%%%%%%%%%%%%%%
In summary we have considered ladder-shaped Rydberg arrays with two or three atoms per site. Originally these simulators were designed with the idea of mimicing closely the evolution of the compact Abelian Higgs model.
We constructed an effective Hamiltonian valid when the size of the rungs is small enough and the distance between the rungs not too small. In all cases, we found that the effective Hamiltonians have the same three types of terms as the target model in Eq.~\eqref{eq:hahm} plus an extra quartic term. The effect of the extra term is significant at positive detuning and is responsible for new quantum phases. More generally, the ladder models have a very rich phase diagram that is currently being explored using QuEra \cite{quera,wurtz2023aquila,zhang2024probing}. Matching the effective theory with the target gauge theory requires $\Delta<0$ (cost for producing electric field). The new phases appear when $\Delta >0$  ($m=\pm1$ form the degenerate ground state). It is possible that a positive detuning could be used as an environment effect relevant in the context of hybrid hadronization \cite{Heitritter:2022jik}. Potential issue with microscopic string breaking generated by extra terms with a large coefficient suggest to study staggered structures (cells with two sites) that could be used to describe models different from the target CAHM. 

It should also be noted that here has been interest in inhomogeneous phases and the Lifshitz regime  for QCD at finite temperature and density \cite{Pisarski:2019cvo}. It has been argued that \cite{Kojo:2009ha} that for massless quarks in the large $N_c$ limit, dimensional reduction occurs 
 and ``chiral spiral” condensates appear \cite{Basar:2008im}. In this scenario, the chiral spiral phase could appear at the end of the crossover line and separate an ordered phase where chiral symmetry is restored from the hadronic (confining) phase where chiral symmetry is broken (see Fig. 6 in \cite{Pisarski:2019cvo}). In this context, exploring the possibility of inhomogeneous phase in simulators is an interesting direction of research. 

 In the future, it would also be interesting to have a simulator where negative (attractive) couplings among atoms could be engineered. This would allow us to cancel the extra terms and to have an effective Hamiltonian identical to the target Hamiltonian. Similar technological needs to be present for quantum simulation of the tricritical Ising model (with differrent types of NNN interactions being negative) \cite{Slagle:2021ene}. As we are moving toward better local control of the individual atoms we should look forward to using new technologies for lattice gauge theory. It would also be interesting to compare the manipulation of the three states associated with a rung in our approach with qutrit simulations \cite{Gustafson:2021qbt,Gustafson:2022xlj} and figure out if the extra quartic term found here could be understood in the context of Symanzik improvement \cite{Carena:2022kpg,Carena:2022hpz}

\begin{acknowledgments}
We thank Sergio Cantu, James Corona, Fangli Liu, Kenny  Heitritter, Steve Mrenna, Shengtao Wang  and members of QuLAT for helpful discussions. 
This work was supported in part by the National Science Foundation (NSF) RAISE-TAQS under Award Number 1839153 (SWT). Computations were performed using the computer clusters and data storage resources of the HPCC, which were funded by grants from NSF (MRI-1429826) and NIH (1S10OD016290-01A1). Y.M. is supported in part by the Dept. of Energy under Award Number DE-SC0019139. J.Z. is supported by NSFC under Grants No. 12304172 and No. 12347101, Chongqing Natural Science Foundation under Grant No. CSTB2023NSCQ-MSX0048, and Fundamental Research Funds for the Central Universities under Projects No. 2023CDJXY-048 and No. 2020CDJQY-Z003. Y. M. thanks the Amazon Web Services and S. Hassinger for facilitating remote access to QuEra through the Amazon Braket.
\end{acknowledgments}

%\bibliography{Spin1Ryd.bib}

%\newpage

\appendix*
%%%%%%%%%%%%%%%%%%%%%%%%%%%%%%%%%%%%%%%%%%%%%%%%%%%%%%%%%%%%%%%
\section{Compact scalar QED}
%%%%%%%%%%%%%%%%%%%%%%%%%%%%%%%%%%%%%%%%%%%%%%%%%%%%%%%%%%%%%%%

The target model in Eq.~\eqref{eq:hahm} is the Lattice scalar QED (sQED) on a $N_s \times N_\tau$  Euclidean spacetime lattice. There are two more columns on two sides, the $0$th and the ($N_s+1$)th columns, controlling the boundary conditions. We use $00$ boundaray condition ($00$BC), where the plaquette (field) quantum numbers on the $0$th and the ($N_s+1$)th columns are zero thus the total charge in the system is zero. In the time continuum limit \cite{prd92,PhysRevD.98.094511}, we obtain the Hamiltonian in the charge representation
\begin{align}
	\label{eq:any-spin-ham-charge}
	\hat{H}_C &= \frac{U}{2}\sum_{1 \leq j, k \leq N_s}  c_{jk} \hat{S}^z_{j} \hat{S}^z_{k} + \frac{Y}{2} \sum_{i=1}^{N_s+1} (\hat{S}_{i}^z)^2 \\ \nonumber
    &- \frac{X}{2} \sum_{i = 1}^{N_s} (\hat{U}_{i}^+ \hat{U}_{i+1}^- + \hat{U}_{i}^- \hat{U}_{i+1}^+) \ , \sum_{i} S_i^z = 0
\end{align}
where $c_{jk} = N_s + 1 - \max\{j, k\}$. The eigenvalues of the operator $\hat{S}^z$ ($\hat{S}^z \ket{n} = n \ket{n}$) are integer charge $n = 0, \pm 1, \pm 2, \ldots$ attached on the vertical links, and the eigenstates define basis of the charge representation. The operator $\hat{U}^{\pm}$ raises (lowers) the charge of a state by one $\hat{U}^{\pm} \ket{n} = \ket{n \pm 1}$. The first term is the self energy of the electric field and is obtained by Gauss's Law. When the  gauge coupling $U = 0$, the spin-$1$ truncation has an infinite-order phase transition from a gapped phase into a BKT critical line \cite{PhysRevB.103.245137}. For nonzero gauge coupling, Eq.~\eqref{eq:any-spin-ham-charge} has unusual long-range interactions, thus it is difficult to design quantum simulators for it. Using Gauss's Law, we can go to the field representation
\begin{align}
\label{eq:any-spin-ham-field}
	\hat{H}_F &= \frac{U}{2}\sum_{p=1}^{N_{s}} \left(\hat{L}^z_{p}\right)^2 
	+ \frac{Y}{2} \sum_{p=1}^{N_s+1} (\hat{L}^z_{p} - \hat{L}^z_{p-1})^2 \\ \nonumber &-
	\frac{X}{2} \sum_{p=1}^{N_s} (\hat{U}^+_{p} + \hat{U}^-_{p})\ ,
\end{align}
where the eigenvalues of $\hat{L}^z$ is the field quantum numbers $m = 0, \pm 1, \pm 2, \ldots$ attached to the plaquettes. There are $N_s$ plaquettes and $N_s+1$ links, $m_0 = m_{N_s+1} = 0$ under 00BC. For open boundary conditions (OBCs), $p$ is taken from $2$ to $N_s$, the Hamiltonian contains multiple charge sectors that have total charge $m_{N_s}-m_1$. The second term in Eq.~\eqref{eq:any-spin-ham-field} can be expanded and then the onsite quadratic term is absorbed into the first term to have Eq.~\eqref{eq:hahm}.

The field representation is identical to the charge representation without applying a truncation. However, the two Hamiltonians are quite different for small spin truncations $|m|, |n|\le S$. For one plaquette with two links, the basis in the field representation is $\ket{-S}$, $\ket{-S+1}$, ..., $\ket{S}$. The basis in the charge representation is $\ket{-S, S}$, $\ket{-S+1, S-1}$, ..., $\ket{S, -S}$. The two Hilbert spaces observe one-to-one mapping. But for two plaquettes with three links, the dimensions of the two Hilbert spaces are not the same, e.g. for $S = 1$, the basis in the field representation is $\ket{-1,-1}$, $\ket{-1,0}$, $\ket{-1,1}$, $\ket{0,-1}$, $\ket{0,0}$, $\ket{0,1}$, $\ket{1,-1}$, $\ket{1,0}$, $\ket{1,1}$, the corresponding states that satisfy Gauss's law in the charge representation are $\ket{-1,0,1}$, $\ket{-1,1,0}$, $\ket{-1,2,-1}$, $\ket{0,-1,1}$, $\ket{0,0,0}$, $\ket{0,1,-1}$, $\ket{1,-2,1}$, $\ket{1,-1,0}$, $\ket{1,0,-1}$. The two states $\ket{-1,2,-1}$, $\ket{1,-2,1}$ are truncated in the charge representation with spin-1 truncation. Generally, for $N_s$ plaquettes with $N_s+1$ links, if the states $\ket{m_1, m_2, ..., m_{N_s}}$ with $|m_p - m_{p+1}| > S$ in the field representation have large energy gap to other states, the effective low-energy basis is a subset of basis in the charge representation with the same spin truncation. 

For spin-1 truncation, we can add a high energy penalty for states $\ket{\ldots, \pm 1, \mp 1, \ldots}$, then we have a new Hamiltonian
\begin{align}
\label{eq:any-spin-ham-field-yp}
\nonumber \hat{H}'_F &= \frac{U}{2}\sum_{p=1}^{N_{s}} \left(\hat{L}^z_{p}\right)^2 
	+ \frac{Y}{2} \sum_{p=1}^{N_s+1} (\hat{L}^z_{p} - \hat{L}^z_{p-1})^2 \\ &-\frac{Y'}{2} \sum_{p=1}^{N_s+1} \hat{L}^z_p \hat{L}^z_{p-1} (\hat{L}^z_{p} - \hat{L}^z_{p-1})^2 -
	\frac{X}{2} \sum_{p=1}^{N_s} (\hat{U}^+_{p} + \hat{U}^-_{p}) \nonumber \\ \nonumber &= (\frac{U}{2}+Y)\sum_{p=1}^{N_{s}} \left(\hat{L}^z_{p}\right)^2
	- (Y+Y') \sum_{p=1}^{N_s+1} \hat{L}^z_{p}\hat{L}^z_{p-1} \\ &+ Y'\sum_{p=1}^{N_s+1} (\hat{L}^z_p)^2 (\hat{L}^z_{p-1})^2 -
	\frac{X}{2} \sum_{p=1}^{N_s} (\hat{U}^+_{p} + \hat{U}^-_{p}),
\end{align}
where the $Y'$ term is only nonzero for states like $\ket{\ldots, \pm 1, \mp 1, \ldots}$. For OBC, the coefficients of $(\hat{L}^z_p)^2$ at $p=1, N_s$ are $U/2+Y/2$ instead of $U/2 + Y$. By setting $Y' \gg 1$, the states like $\ket{\ldots, \pm 1, \mp 1, \ldots}$ in the field representation that have no counterpart in the charge representation will be suppressed. When $N_s=2$, the two representations have the same Hilbert space if states $\ket{\pm 1, \mp 1}$ are suppressed. The energy gaps for the two representations should be the same in the large $Y'$ limit. For $N_s = 2, U=0, Y=X=1$, the energy gap for $\hat{H}_C$ is $\Delta E = 1$, while $\Delta E \approx 1.44$ for $\hat{H}_F$. If we set $Y' = 1000$, $\Delta E = 1.00008$ for $\hat{H}'_F$. For $N_s \ge 3$, the Hilbert space for the field representation in the large $Y'$ limit is a subset of that for the charge representation and the energy gap cannot be equal. But the energy gap in the field representation can be closer to that in the charge representation for large $Y'$. For example, $N_s = 10, U=0, Y=X=1, Y'=1000$, the energy gaps for $\hat{H}_C$, $\hat{H}_F$, and $\hat{H}'_F$ are $0.357$, $0.605$, and $0.532$, respectively.

Since the states in the charge representation like $\ket{\ldots, 111, \ldots}$ have no correspondence in the basis of the field representation, the field representation with spin-1 truncation has no infinite-order phase transition that exists in the charge representation with spin-1 truncation even if $Y'\gg 1$ and $\ket{\ldots, \pm 1, \mp 1, \ldots}$ state are suppressed. When $S$ increases to $\infty$ in the spin-$S$ truncation, the difference between the spectrum of Eq.~\eqref{eq:any-spin-ham-charge} and that of Eq.~\eqref{eq:any-spin-ham-field} diminishes to zero \cite{PhysRevD.98.094511,Zhang:2018ufj}. For finite $S$, adding high energy penalty like the $Y'$ term in the spin-1 truncation to suppress unwanted states can increase the precision of quantum simulation proposals using the field representation. The spin-1 truncated model considered here can have other interesting critical properties and we are interested in quantum simulation of Eq.~\eqref{eq:any-spin-ham-field-yp} using Rydberg ladders. Our discussion in the main text reveals that the effective spin-1 Hamiltonian for the 2- or 3-leg Rydberg ladder in Eq.~\eqref{eq:effectivehamgeneral} has exactly the same form as $\hat{H}'_F$ in Eq.~\eqref{eq:any-spin-ham-field-yp} with $D = U/2+Y$, $R = Y+Y'$, $R'=Y'$, and $J=X/2$.

We have discussed the matching between the target model in Eq.~\eqref{eq:hahm} and the effective Hamiltonian for the two-leg Rydberg ladder in Eq.~\eqref{eq:effectiveham2legNN}. If we match Eq.~\eqref{eq:any-spin-ham-field-yp} with Eq.~\eqref{eq:effectiveham2legNN}, $Y$ should be negative because $0 < R < R'$ in Eq.~\eqref{eq:effectiveham2legNN}, while $Y$ is positive in the original Hamiltonian formulation of sQED. By matching the effective Hamiltonian for the three-leg Rydberg ladder in Eq.~\eqref{eq:effectiveham3legNN} to the target model Eq.~\eqref{eq:hahm}, $\hat{H}^{\rm{eff}}_{\rm{3LR}} = \hat{H}_{CAHM} + \rm{Const.} \hat{I}$ ($\rm{Const.}$ is a constant), we have the following equations
\begin{eqnarray}
\label{eq:matcho3legobc}
\nonumber D &=& \Delta_0 + \Omega^2(B-A)/4 + 2(V_2-V_1) \\ \nonumber 
Y &=& R = (V_1-V_3)/2 \\ \nonumber 
X &=& 2J = \Omega^2\Gamma/2 \\
R' &=& \left(3V_1 + V_3\right)/2 - 2V_2 = 0
\end{eqnarray}
and $\rm{Const.} = -(\Delta+\Delta_0)N_s - N_s\Omega^2B/4 + V_1(N_s-1)$. Notice that the coupling of $(\hat{L}^z_i)^2$ on the boundary is different from that in the bulk, which can be remedied by tuning the local detuning on boundary sites. The problem here is that $4V_2 = 3V_1 + V_3$ cannot be satisfied for any aspect ratio, but $R/R'$ is a monotonically increasing function of $a_x/a_y$ and $R/R' > 1$ for $a_x/a_y > 2.25$. We can tune the aspect ratio such that the effect of the quartic term is small. Since $R/R' > 1$ is realizable, $\hat{H}^{\rm{eff}}_{\rm{3LR}}$ in Eq.~\eqref{eq:effectiveham3legNN} can be matched to $\hat{H}'_F$ in Eq.~\eqref{eq:any-spin-ham-field-yp}. One can see that the three-leg Rydberg ladder is a more programmable quantum simulator for spin-1 models. Although the spin-1 sector is not the lowest band in real Rydberg systems, one can first prepare a product state like $\ket{\ldots,0,0,0,\ldots}$ in the spin-1 sector, and then adiabatically prepare the lowest energy state for certain parameter regime in the energy band of the spin-1 sector.

\end{document}